\documentclass[12pt,preprint]{aastex}

\slugcomment{ver. 07 Feb. 2005}
\shorttitle{Dust in the Photospheric Environment III. }
\shortauthors{T. Tsuji}

\begin{document}

\title{ DUST IN THE PHOTOSPHERIC ENVIRONMENT III. A FUNDAMENTAL ELEMENT IN 
THE CHARACTERIZATION  OF ULTRACOOL DWARFS}

\author{TAKASHI TSUJI}
\affil{Institute of Astronomy, School of Science, The University of Tokyo \\
2-21-1, Osawa, Mitaka, Tokyo, 181-0015, Japan}
\email{ttsuji@ioa.s.u-tokyo.ac.jp}

\begin{abstract}

Recent photometry of L and T dwarfs revealed that the infrared colors show 
a large variation at a given effective temperature and,
within the framework of our Unified Cloudy Model (UCM),
this result can be interpreted as due to a variation of the critical 
temperature ($T_{\rm cr}$) which is essentially a measure of the thickness 
of the dust cloud. 
Especially, the variation is the largest at $T_{\rm eff} \approx 1400 \pm 
100$\,K, where the transition from L to T types takes place. Thus the L/T 
transition is associated with the drastic change of the thickness of the dust 
cloud at $T_{\rm eff} \approx 1400$\,K, but the reason for this change is 
unknown. Once we allow $T_{\rm cr}$ to vary at given $T_{\rm eff}$ and 
log\,$g$ in our UCMs,
the two-color diagram and color-magnitude diagram can be well 
explained as the effect of   $T_{\rm eff}$, log\,$g$, and $T_{\rm cr}$,
but not by the effect of $T_{\rm eff}$  and log\,$g$ alone.  
This result implies that $T_{\rm cr}$ will be one of the 
important parameters needed for characterization of dusty dwarfs. 
The effects of $T_{\rm eff}$ and  $T_{\rm cr}$ on individual  spectra, 
however,  are difficult to discriminate, since changing of $T_{\rm eff}$ at 
a fixed $T_{\rm cr}$ on one hand and changing of $T_{\rm cr}$ at a fixed 
$T_{\rm eff}$ on the other essentially have the same effect on the spectra.
We show that the degeneracy of  $T_{\rm eff}$ and  $T_{\rm cr}$ can be 
removed to some extent by the analysis of the spectral energy distribution 
(SED) on an absolute scale. The reanalysis of a selected sample of spectra 
revealed that the L-T spectral sequence may not necessarily be a sequence of 
$T_{\rm eff}$, but may reflect a change of the thickness of the dust cloud, 
represented by $T_{\rm cr}$ in our UCM.
Although this unexpected result is based on a limited sample,
an odd ``brightening'' of the absolute $J$ magnitudes plotted against the 
L-T spectral types may also be a manifestation that the L-T spectral sequence
is not necessarily a temperature sequence. This is because $M_{\rm bol}$ 
based on the same photometry data also shows a similar ``brightening''. 
Then, the ``$J$-brightening'' may not be due to any atmospheric effect and
hence should not be a problem to be solved by model atmospheres including
the UCMs.  Thus, almost all the available observed data are reasonably well 
interpreted with the UCMs, whose full capability emerges once  $T_{\rm cr}$
is introduced as the fifth parameter in addition to the usual four parameters
(i.e., chemical composition,  $T_{\rm eff}$, log\,$g$, and micro-turbulent 
velocity) needed to characterize stellar spectra in general.
  
\end{abstract}

\keywords{molecular processes --- stars: atmospheres --- stars: fundamental 
parameters ---stars: late-type --- stars: low-mass, brown dwarfs ---   }

\section{INTRODUCTION}

In any branch of natural science, classification of objects to study is a 
fundamental step, from which next developments  will be conceived and 
undertaken. Thus, it is quite natural that the spectral classification of 
ultracool dwarfs including brown dwarfs has been initiated as soon as some 
dozens of dwarfs cooler than  type M have been discovered in the late 1990's.
A new spectral class L was assigned to these objects, whose prototype is
GD165b \citep{bec88},
and  spectral subclasses were defined by considering various spectral features 
in the optical spectra \citep[][]{kir99,mar99}. Soon the near infrared
spectra are used for the spectral classification \citep[][]{rei01,tes01}. 
It did not take long before a considerable number of cool brown dwarfs, whose 
prototype is Gl\,229b \citep{nak95}, have been discovered, and an attempt to 
classify them under a new spectral class T was undertaken \citep[][]{bur02a}.
Finally, a unified classification scheme of L and T types was proposed 
with the use of the near infrared spectra \citep[][]{geb02}. 
Now, a question is what the L and T types mean.

The stellar spectral classification, which has the present form 
at the beginning of the 20th century, was done on purely empirical basis
at first. It was later interpreted as a temperature 
sequence and further developed to the two dimensional  
classification including the luminosity effect.
The physical basis of the present two dimensional stellar spectral 
classification such as MK system is quite clear:
The basic stellar properties represented by luminosity $L$ and radius $R$,
 from which $T_{\rm eff}$ and log\,$g$ are defined, can be
determined uniquely as a function of time once the initial mass and
chemical composition are given (Vogt-Russell theorem; e.g. Chandrasekhar 
1939). Thus any object on  the HR diagram has its corresponding spectral type.
Later it was recognized that the characterization of stellar spectra 
requires an additional parameter referred to as the micro-turbulent velocity 
\citep{str34} which, however, is more difficult to predict for individual
objects. 

A problem is if the case of L and T types can be a simple extension of this 
paradigm. A new feature of L and T dwarfs is that dust forms in their
photosphere, and we proposed that the L-T spectral sequence can be interpreted 
with a model which we referred to as the Unified Cloudy Model 
\citep[][]{tsu01}: In this UCM, dust forms at the condensation temperature
$ T_{\rm cond}$,  but soon grows to be too large at a slightly lower
temperature which was referred to as the critical temperature 
 $ T_{\rm cr}$, and segregates from the gaseous mixtures.
Thus a dust cloud forms in  the layer where 
$  T_{\rm cr} \la T \la T_{\rm cond}$. In this model,
$ T_{\rm cond}$ is essentially determined by the thermodynamical data
and thus well predictable for the basic parameters such as  $ T_{\rm eff}$, 
log\,$g$, and metallicity. At first, we also thought that
$ T_{\rm cr} $ may in principle have a causal relation to the
basic parameters even if we do not yet know the exact relation.
Then, we assumed a constant value for $ T_{\rm cr} $
for simplicity, and estimated it empirically. Since $ T_{\rm cr} $ is
a measure of the thickness of the dust cloud, which has a direct effect on the
infrared colors, we estimated  $ T_{\rm cr}$ to be about  1800\,K from
the red limit of $J-K$ (Tsuji 2002; hereafter be referred to as Paper I).
Then it was possible to interpret the spectral sequence from L to T as a 
temperature sequence (Tsuji, Nakajima, \& Yanagisawa 2004;  hereafter be 
referred to as Paper II).

Recent progress in observations, however, casts some doubts as to whether
if the L-T spectral types are simple extension of the stellar spectral types
to the cooler temperatures.  Especially, the effective temperatures based on
the bolometric luminosities determined from the recent astrometry 
\citep[][]{vrb04} and photometry extended to $L^{'}$ and $M^{'}$ bands
\citep[][]{gol04} appeared to be nearly constant between middle L and early T 
types \citep[][]{vrb04, gol04}. This fact suggests that the L-T
spectral sequence may  not necessarily be a temperature sequence   
and that the spectral sequence is controlled by parameter(s) other than 
$ T_{\rm eff}$. Also, it appears that the infrared colors extended to
a large sample of L and T dwarfs \citep[][]{kna04} show a large variation 
especially if  they are plotted against $ T_{\rm eff}$.
This fact implies that $ T_{\rm cr}$, which is directly related to the
infrared colors as noted above, cannot be a constant throughout L and T 
dwarfs as assumed in our Papers I and II, but should show large changes 
independently of the basic stellar properties such as $ T_{\rm eff}$ and 
log\,$g$ (see Sect.2.1 as for details).  
Hence any conclusion based on the assumption of a constant $ T_{\rm cr}$
must be reconsidered. 

In our UCMs, the property of the dust cloud is represented by a
single parameter, $ T_{\rm cr}$, which is a measure of the thickness
of the dust cloud. This is essentially equivalent to have specified the
dust column density in the cloud for the distribution of dust grains
given by the static model photosphere. The cloud consists of grains
smaller than the critical radius (Paper I) and the grain size is assumed to be
small enough to be in the Rayleigh regime (i.e. $r_{\rm gr} << \lambda $).
No other details of the cloud are specified in the present UCMs. 
If we are to specify some details of the clouds,
we must make some, possibly ad hoc, assumptions, since we do not know  
the details how clouds are formed in the photospheres of cool dwarfs
at present. It is to be remembered that $ T_{\rm cr}$ is an empirical
parameter and it is not derived from any particular model of cloud 
formation.   The new feature that the thickness of the dust cloud changes 
unpredictably implies that the cloud formation may take place under 
chaotic condition and/or may already be a meteorological phenomenon.
Since theoretical prediction of an accidental event is more difficult, it may 
be useful to have emphasized an empirical approach 
before more serious effort of detailed modeling of the cloud formation
will be done.  It is to be remembered that a purpose of modeling is to
describe complicated astronomical phenomenon by a simple picture as 
far as possible at first, and our UCM is presently at this stage.  

In this paper, we hope to examine to what extent our simple model could 
explain the available observed data on L and T dwarfs.
First, we will show that most  observations such as colors and 
magnitudes (Sect.2) as well as spectra (Sect.3) can be accounted for 
reasonably well by our UCMs 
if we assume that the critical temperature $T_{\rm cr}$ varies  between the 
surface temperature ($T_{\rm 0}$) and the condensation temperature 
($T_{\rm cond}$) at fixed $T_{\rm eff}$ and log\,$g$.
This fact confirms that the thickness of the dust cloud, or  $T_{\rm cr}$
in our UCM, is an important parameter in interpreting the observed data of 
dusty dwarfs. Next, we discuss how the basic properties including effective 
temperature, luminosity, and spectral classification can be interpreted by
the applications of the UCMs with the variable $T_{\rm cr}$ (Sect.4). However, 
some conclusions appear to be quite unexpected:
For example, it is shown that the L-T spectral sequence is not necessarily
a sequence of $T_{\rm eff}$ but $T_{\rm cr}$ plays dominant role
at least between middle L and early T types. 
This is because both of $T_{\rm eff}$ and $T_{\rm cr}$ have
significant effect on the dust column density in the observable photosphere,
and $ T_{\rm cr}$ is sometimes more important than  $T_{\rm eff}$
in characterization of the photosphere of dusty dwarfs. In our formulation
of the UCMs, this new parameter
$T_{\rm cr}$ distinguishes dusty dwarfs from other stars, which are 
characterized by four parameters, namely, chemical composition, $T_{\rm eff}$,
log\,$g$, and micro-turbulent velocity.

\section{Colors and Magnitudes}

Recent infrared photometry revealed that infrared colors plotted against
$T_{\rm eff}$ show a large variation, and this fact implies that
the thickness of the dust cloud should be changing independently 
of $T_{\rm eff}$. 
In our UCMs, the critical temperature $T_{\rm cr}$ is introduced as a
 measure of the thickness of the dust cloud, but it was assumed that
$T_{\rm cr}$ is constant throughout L and T dwarfs for simplicity in Papers I 
and II. This is partly because the scatters of the infrared colors plotted
against the spectral type appeared to be not so large as those plotted against
$T_{\rm eff}$. In view of the new feature, however, we think it
an over-simplification to have assumed a unique value for $T_{\rm cr}$.
Furthermore, this assumption might have  
suppressed the potentiality that the UCMs could be able to have if they 
are applied more properly.  We  now regard $T_{\rm cr}$ as a free parameter
to be variable between the condensation and surface temperatures. Then,
we demonstrate that such purely empirical data as the two-colors diagram
and color-magnitude (CM) diagram of L and T dwarfs can be explained with
the three parameters, $T_{\rm eff}$, log\,$g$, and $T_{\rm cr}$ while
not at all by the two parameters, $T_{\rm eff}$ and log\,$g$.

\subsection{Infrared Colors and the Critical Temperature}

Infrared colors plotted against L - T spectral types first show reddening
from early to late L types and, after a reversal at late L, show bluing  
from early to late T types \citep[e.g.][]{leg02}. Recent progress in
observations revealed some new features:   
First, thanks to the extended parallax data \citep[][]{vrb04} and 
bolometric corrections based on the $L^{'}$ and $M^{'}$ photometry
\citep[][]{gol04}, absolute bolometric luminosities are obtained for
a large sample of L and T dwarfs. Then empirical effective temperatures
can be derived if the radii of L and T dwarfs can be known. For this purpose, 
\citet[][]{vrb04} assumed a constant radius of 0.9$R_{\rm Jupiter}$ while
\citet[][]{gol04} applied the radii based on evolutionary models.
The resulting values of $T_{\rm eff}$ at 3\,Gyrs by \citet[][]{gol04}
mostly agree with those by \citet[][]{vrb04} (see Sect.4.2),   
and this is what can be expected since the parallax
data and bolometric corrections are mostly the same in the two works.
We plotted the observed colors on the MKO system \citep[][]{kna04} against  
the empirical values of $T_{\rm eff}$ \citep[][]{vrb04}
rather than against the L - T spectral types in Fig.\,1, where $T_{\rm eff}$ 
values based on the bolometric luminosities are shown
by the large circles and those based on the estimations with the 
$T_{\rm eff}$ - Sp.\,Type relation by the small circles. In Fig.\,1, L
and T dwarfs are shown by the filled and open circles, respectively, and
it appears that the transition from L to T types takes place at
$T_{\rm eff} \approx 1400 \pm 100$\,K.  It is remarkable
in Fig.\,1 that the scattering is very larger especially at around 
$T_{\rm eff} \approx 1400$\,K.  A preliminary result that the infrared 
colors such as $J-K$ plotted against $T_{\rm eff}$ (and not against the 
L - T types) show a drastic change at the L - T transition was already
noted by us \citep[][]{tsu05}, independently by Marley, Cushing, \& Saumon 
(2005), and  also by \citet{leg05}.

Although it was known that the scatters of the infrared colors plots against 
the spectral type are rather large in L dwarfs (e.g. Fig.\,3 of 
Knapp et al. 2004), the plots of the same colors against effective temperature
revealed that the scatters are much larger especially at around the L -  T 
transition. It is remarkable that the plots of the same infrared colors
against the spectral type and those against the effective temperature
are so different. This means that an implicit assumption that the L - T 
spectral sequence is a temperature sequence may not necessarily be correct, 
even though this might be a rather natural assumption in view of
the smooth change of the infrared colors plotted against the spectral types.
In fact, this conventional assumption now appears to be not consistent
with the recent empirical effective temperature determinations \cite[e.g.][]
{gol04,vrb04}.
 
The modest scatters of the infrared colors plotted against the L-T spectral 
were already interpreted as due to variations in the altitudes, spatial 
distributions, and thicknesses of the clouds \citep[][]{kna04}. 
The scatter of the near infrared colors at the same L spectral types was 
previously noted on the 2MASS data by \citet[][]{ste03} and interpreted as 
due to the change of the cloud opacity which depends on the sedimentation 
efficiency $f_{\rm sed}$ of the cloud model of \citet[][]{ack01}. 
In our UCM, the thickness of the dust cloud is represented by a parameter
referred to as the critical temperature $T_{\rm cr}$ (Tsuji 2002), but
we assumed it to be a constant for simplicity in our previous applications
of UCMs \citep[e.g.][]{tsu03,tsu04}. This assumption, however, should be
reconsidered in view of the much larger scatter
of the infrared colors as revealed in Fig.\,1.

For this purpose, we first examine the effect of gravity and plot the 
predicted infrared colors based on the UCMs with
log\,$g$ = 4.5, 5.0 and 5.5 in Fig.\,1a ($T_{\rm cr} = 1800$\,K throughout). 
We applied the filter response functions (Tokunaga, Simon, \& 
Vaccatok 2002) to the predicted fluxes based on the case I 
(band model) opacities for methane (see Paper II as for detail).
Inspection of Fig.\,1a reveals that some scatters of the observed infrared 
colors can be explained as the gravity effect (e.g. $J-K$ at $T_{\rm eff} 
\ga 1600$\,K and $H-K$ at $T_{\rm eff} \la 1300$\,K) and, for this reason,
we previously thought that the scatter of $J-K$ can be explained by the 
gravity effect (Paper II).  Such a gravity effect on the infrared colors
was also discussed by \citet{kna04} on the basis of the cloudy models by
\citet{mar02}. However, it is clear in Fig.\,1a that the large variations of 
the infrared colors near the L - T transition
cannot be explained by the gravity effect. It is to be
noted that the large variations of the infrared colors at $T_{\rm eff} 
\approx 1400$\,K has not been recognized before, and it is the smaller 
variations of the infrared colors plotted against the L - T spectral types 
that were thought to be interpreted as the gravity effect in our Paper II.
Now the situation is completely different because of the drastic changes
of the infrared colors at $T_{\rm eff} \approx 1400$\,K, and we must now
recognize that the large scatter of the infrared colors near the L - T 
transition cannot be explained by the gravity effect. 

Next, we overlaid the predicted colors based on the UCMs with $T_{\rm cr}$ = 
1700, 1800, 1900K, and $T_{\rm cond}$ in Fig.\,1b (log\,$g$ = 5.0 throughout).
 Previously, we interpreted that the scatters of the colors at their
red limit in late L dwarfs can roughly be explained  by
$T_{\rm cr} = 1800 \pm 100$\,K.
For this reason, we assumed that   $T_{\rm cr} \approx 1800$K may apply to 
all the cool dwarfs from L to T throughout in our previous applications of 
the UCMs (Papers I and II).  However,  this is clearly an over-simplification 
in view of the larger variation around  the L - T transition at $T_{\rm eff}
\approx 1400$,K (Fig.\,1), which has not been recognized before when the
infrared colors were plotted against the L - T spectral type. Now, to explain 
the large variations of the infrared colors near the L - T transition, we 
must assume from Fig.\,1b that $T_{\rm cr}$ in our UCM, or the thickness of 
the dust cloud, should be different at a given $T_{\rm eff}$
\footnote{It is to be noted that the thickness of the cloud, or more properly,
the dust column density in the observable photosphere, is not necessarily the 
same even if $T_{\rm cr}$ is fixed, since $T_{\rm cond}$ shows continuous 
changes through L to T dwarfs (see Fig.\,3 in Paper I). In fact, this may be
the reason why infrared colors, largely controlled by the dust extinction,
change with $T_{\rm eff}$. However, the change of the dust column density
due to $T_{\rm cr}$ itself introduced here is independent of $T_{\rm eff}$ 
and of different nature.}. Moreover, the variation of the thickness of the
dust cloud should be quite drastic such that $T_{\rm cr}$ changes from
$T_{\rm cond}$ to below 1700\,K in our UCM. Note that the 
case of $T_{\rm cr} \approx T_{\rm cond}$ implies that the dust cloud 
effectively disappears.

In the UCMs, $T_{\rm cond}$ is predictable based on the thermochemical
data and hence it is well defined for given $T_{\rm eff}$ and log\,$g$. In 
general, $T_{\rm cond}$ is higher at at higher gravities
and/or at lower effective temperatures, because of the higher gas pressures 
in these cases. For example,  $T_{\rm cond}$ values of iron are 1900 and 
2200\,K for models of $T_{\rm eff} = 1800$ and
1000\,K, respectively (log g= 5.0 throughout; see Fig.\,3 in Paper I). We have 
already considered a possibility of variable $ T_{\rm cr}$ but in such a way
that $ T_{\rm cr}$ depends on $T_{\rm eff}$ and log\,$g$. In fact, the 
thickness of the cloud will be very thin in early L dwarfs for 
$ T_{\rm cr} \approx 1800$\,K, since $ T_{\rm cond}$ will already be close to
1800\,K for the models of relatively high $ T_{\rm eff}$, and we considered a 
possibility that $ T_{\rm cr}$ may be lower for the models of high 
$ T_{\rm eff}$ (Paper I)\footnote{In the early L dwarfs, 
observed colors appear to be  redder than the predicted ones for
any value of $ T_{\rm cr}$ in Fig.\,1. This result may not necessarily
be related to dust, but may be due to some other problems such as of the  
gaseous opacities.}. With such a possibility in mind, we have prepared 
the UCMs with several values of $ T_{\rm cr}$. However, the fact realized in
nature turns out to be beyond what we have imagined: The variation of 
$T_{\rm cr}$ appears to be independent of $T_{\rm eff}$ and log\,$g$ 
(Figs.\,1a \& b), and to be quite large, but 
the origin of the variation cannot be understood at all.

The fact that the transition from L to T takes place at  $T_{\rm eff} 
\approx 1400$\,K implies that $T_{\rm eff}$ is an important factor in the 
L - T transition, but it is also clear that 
another effect plays decisive role. Within the framework of our UCMs, this 
effect is identified with
$T_{\rm cr}$ which is a measure of the thickness of the dust cloud.
It is as if something might have happened at $T_{\rm eff} \approx 1400$\,K
to induce the change of the thickness of the cloud  in such a way that L 
dwarfs with the 
lower values of $T_{\rm cr}$ evolve to T dwarfs with the higher values of 
$T_{\rm cr}$. 
The rapid and almost discontinuous change of the infrared colors at the 
L - T transition in a very small $T_{\rm eff}$ range is also clearly  
illustrated in Fig.\,5 of \citet[][]{mar05}. Since the L - T 
transition is usually discussed in connection with the CM diagrams, we will 
return to this subject in Sect. 2.3.

A further complication is that $T_{\rm cr}$ and  $T_{\rm eff}$ have nearly
the same  effect on the infrared colors. 
For example, we interpreted the reddening of the infrared colors in
L dwarfs as due to the increase of the dust column density in the
observable photosphere with the decreasing $T_{\rm eff}$. But
the dust column density will also increase with decreasing $T_{\rm cr}$
even at the fixed $T_{\rm eff}$. Thus, $T_{\rm eff}$ and $T_{\rm cr}$
have the same effect on the infrared colors. In Fig.\,1b, it was
possible to separate the effect of $T_{\rm eff}$ and $T_{\rm cr}$, and
this is  because $T_{\rm eff}$ was
known by other method in this case. Generally, it is difficult to
remove the degeneracy of $T_{\rm eff}$ and $T_{\rm cr}$ unless an
independent information on one of the two can be available. This problem
will further be discussed in Sect.3.

\subsection{Two-Colors Diagram}

We plotted the observed $J-H$ against $H-K$ on the MKO system \citep[][]{kna04}
in Fig.\,2a, where  L and T dwarfs are shown by filled and open circles,
respectively. We overlaid the predicted colors 
based on the UCMs with the three
values of log\,$g$ = 4.5, 5.0 and 5.5, but with the constant value of
$T_{\rm cr} = 1800$\,K throughout.  The effect of gravity is minor
in  L and early T dwarfs, and the observed scatters of the colors
cannot be explained as the gravity effect for these dwarfs. On the
other hand, the scatters of the observed colors in middle and late T dwarfs
are well explained as the gravity effect. Thus we agree as to the gravity 
effect in T dwarfs with \citet[][]{kna04} who applied the models
of \citet{mar02}. Note that our UCMs cover 
the range of $T_{\rm eff}$ between 700 and 2600\,K. Thus an object at the 
lower left corner 2MASS\,J0937+2931 (T6p) may have log\,$g$ somewhat lower 
than 5.0 and $T_{\rm eff}$ somewhat lower than 700K from the colors alone 
if metallicity is normal. 

In Fig.\,2b, we overlaid the predicted colors based on the UCMs with 
 $T_{\rm cr} = T_{\rm 0}$ (case B), 1700, 1800, 1900\,K and $T_{\rm cond}$ 
(case C) on the same observed data as in  Fig.\,2a (log\,$g$ = 5.0 
throughout).  It is confirmed that most of the colors of L and early 
T dwarfs can be explained by changing the $T_{\rm cr}$ values from 
$T_{\rm cond}$ (case C) to $T_{\rm 0}$ (case B). 
Several early T dwarfs are found in the region of 1900\,K$ < T_{\rm cr} < 
T_{\rm cond} $ and it is interesting if this fact implies that the dust 
cloud is really very thin in the early T dwarfs.  Few objects are found 
on the right side of the extreme case of fully dusty models (case B).
The effect of $T_{\rm cr}$ is very small in middle and 
late T dwarfs and the scatters of the observed colors are mostly explained 
as the gravity effect as noted already.

To summarize,  almost all the data points in the observed $(J-H, H-K)$ 
diagram could
be explained by changing the three parameters $T_{\rm eff}$, log\,$g$, and
$T_{\rm cr}$ in appropriate ranges in our UCMs.  It should be noted that
the role of $T_{\rm cr}$ is essential (Fig.\,2b) while $T_{\rm eff}$ and
log\,$g$ alone could not explain the two-colors diagram (Fig.\,2a).

\subsection{Color-Magnitude Diagram}

Recent progress of astrometry and photometry of ultracool stars finally
made it possible to compare CM diagrams with the 
theoretical evolutionary tracks of substellar objects. Our previous attempt 
on the $(M_{J}, J-K)$ diagram \citep[][]{tsu03}, however, was not fully 
successful, but received some criticisms. First, some early T dwarfs that 
show the so-called ``$J$-brightening'' was interpreted as the very young 
low-mass objects, but such a hypothesis does not necessarily find 
observational support. The so-called ``$J$-brightening'' seen on the plot of
$M_{J}$ vs. Sp.\,Types may be  an artifact of the L-T spectral sequence
not representing a temperature sequence as will be shown in Sect.4.1, but  
the ``$J$-brightening'' on the CM diagram still remains to be explained. 
Second, some late L dwarfs are too faint to be explained by our
previous analysis.

We must point out that our previous attempt was based on an assumption
of a uniform value of $T_{\rm cr} \approx 1800$K throughout the L-T
dwarfs, but this assumption can no longer be supported as noted in 
Sects.2.1 and 2.2. We then examine the effect of changing $T_{\rm cr}$ on 
the CM diagram for the three evolutionary tracks of $M = 10, 42,$ and 
70\,$M_{\rm Jupiter}$ \citep[][]{burr97} separately, and compared the 
results with the observed data on the MKO system \citep[][]{kna04}. 
We evaluate $J-K$ and $BC_{\rm J}$ based on the UCMs with  different values 
of $T_{\rm cr}$
\footnote{Some examples of the values of $BC_{\rm J}$ (CIT system) for 
several values of $T_{\rm cr}$ are given in Fig.\,2 of \citet[][]{tsu03}.},
using the filter response functions of the MKO system \citep[][]{tok02}, 
and  converted ($T_{\rm eff}$, $M_{\rm bol}$) to 
$(J-K, M_{J})$ with the use of these results. First, the case of $M = 
10\,M_{\rm Jupiter}$ are shown in Fig.\,3a for $T_{\rm cr} = 1700, 1800,$ 
1900\,K  and $T_{\rm cond}$ (case C) .
The agreement between observed and predicted loci is generally poor except
for early and middle L dwarfs. This fact can be interpreted that  
the  low-mass brown dwarfs are more difficult to observe because of
their short lifetimes and thus they may be under-sampled in the presently 
observed  sample of brown dwarfs.

The same analysis is done for the case of $M = 42\,M_{\rm Jupiter}$, and
the results are shown in Fig.\,3b. In this case, most of the observed
data points, except for late L dwarfs, are covered by the predicted tracks 
with the values of $T_{\rm cr}$ between 1700\,K and $T_{\rm cond}$.   
The so-called ``$J$-brightening'' of some early T dwarfs are now explained 
by the models with very thin clouds or with no cloud as the limiting case of 
reduced cloud thickness. This result is consistent with Fig.\,1b suggesting 
that $T_{\rm cr}$ should be quite high in some early T dwarfs, and
with the suggestion from the two-colors diagram discussed in Sect.2.2 that 
the thickness of the cloud may be quite thin (i.e., $T_{\rm cr} > 1900$\,K)
in early T dwarfs. Finally, the case of $M = 70\,M_{\rm Jupiter}$ is shown 
in Fig.\,3c, and now the very low luminosities of some late L dwarfs are 
explained by the models with the thick cloud (i.e. $T_{\rm cr} \approx 
1700$\,K).As to the very faint L dwarfs slightly below the predicted track for
$T_{\rm cr} = 1700$\,K, possibilities of the lower $T_{\rm cr}$ and/or higher 
masses may be considered.

Summarizing, the observed ($M_{J}, J-K)$ diagram can reasonable be well
reproduced with the evolutionary tracks by \citet[][]{burr97} 
\footnote{Also, the use of the isochrones by \citet[][]{cha00} will result 
in the similar conclusion as can be inferred from our previous result 
\citep{tsu03}.} and the UCMs in which the  parameter $T_{\rm cr}$ is allowed
to change by several hundreds Kelvin. The rapid bluing and brightening
at the transition from L to T can be explained primarily by the immersion of 
the dust cloud from the optically thin to thick region at about $T_{\rm eff} 
\approx 1400 \pm 100$\,K but, at the same time, the thickness of the dust
cloud should decrease  as suggested by the increase of $T_{\rm cr}$ (Fig.\,1b).
Such a change of the thickness of the dust cloud cannot be explained by the 
known theories of the structure and evolution of brown dwarfs, and probably
not yet identified  dynamical process must be called for.

In a recent paper, \citet{kna04} pointed out several difficulties
in  our previous model of the L - T transition \citep{tsu03}, but we hope 
that these issues are mostly resolved by our present version of UCMs with the 
variable cloud thickness.  It is to be noted, however, that our UCMs assume 
a homogeneous cloud throughout. A possibility of  clouds with holes 
has been proposed to explain the rapid transition from L to T as well as 
the ``$J$-brightening'' by \citet[][]{bur02b}, and further discussed by  
\citet[][]{kna04} who referred to it as the ``patchy clouds'' model. 
We think that such an inhomogeneity does not necessarily 
be required  at present, even though such a possibility may not be excluded
a priori. Also, \cite{kna04} suggested another possibility referred to as 
the ``sudden downpour'' model based on the cloudy models of \citet{mar02}:
In this model, L dwarfs first cools at essentially constant sedimentation
efficiency $f_{\rm sed}$ and then $f_{\rm sed}$ begins  to  increase from 
$\approx 3$ to infinity at around $T_{\rm eff}  \approx 1300$\,K. This
rapid increase in  $f_{\rm sed}$ will bring
rapid cloud thinning and hence rapid bluing infrared colors. 
We think that this model is more acceptable than the ``patchy clouds'' model.
In fact, the``sudden downpour'' model and  our UCM with variable cloud 
thickness are essentially based on the same concept  in that both treatments
are to specify the cloud properties  with simple model parameters, 
the sedimentation efficiency $f_{\rm sed}$ and the critical temperature 
$T_{\rm cr}$, respectively.  Anyhow, all these models including ``sudden 
downpour'' model, ``patchy clouds'' model, and our UCM with variable cloud
thickness agree that the nature of the clouds shows drastic changes 
during the L - T transition at $T_{\rm eff} \approx 1400 \pm 100$\,K, and
to identify the mechanism of driving such changes  should
be a key to realize the transition from L to T types.

\section{Infrared Spectra} 

In general, stellar spectra can be characterized by the four parameters, 
namely, chemical composition, $T_{\rm eff}$, log\,$g$, and micro-turbulent 
velocity $v_{\rm micro}$.  In the case of dusty dwarfs, however,
an additional parameter that specify the nature of the dust cloud had to be 
specified for properly interpreting the infrared colors and magnitudes 
(Sect.2) and we introduced  $T_{\rm cr}$ for this purpose.
In this paper, we assume the standard composition throughout and metallicity 
effect will not be considered for simplicity. Also, we assume $v_{\rm micro} = 
1$\,km s$^{-1}$ throughout. Then, the effectively important parameters 
are $T_{\rm eff}$, log\,$g$, and $T_{\rm cr}$,  and a problem is how to 
interpret the spectra of individual objects characterized by the three 
parameters. Our UCMs are already  prepared for such applications in that 
they are formulated with the critical temperature  $T_{\rm cr}$ as an 
important ingredient. The introduction of the new free parameter opens 
a possibility of more flexible analyses of the observed spectra. Our
reanalysis reveals that the previous interpretation of the spectral sequence 
of ultracool dwarfs such as given in Paper II was not correct, and we
believe that we arrive at a more correct interpretation in this paper.

\subsection{Degeneracy of the Effective and Critical Temperatures }

With a proper calibration, spectrum observed with linear detectors
can be regarded as a spectral energy distribution (SED) apart from
an absolute scale. Such a relative SED is usually analyzed by means of the
predicted SEDs based on models, but there is a problem as follows
in the case of dusty dwarfs:  The  dust column density in the observable 
photosphere depend on $T_{\rm eff}$ through its effect on $T_{\rm cond}$ on 
one hand and on $T_{\rm cr}$ through its direct effect on the thickness of 
the cloud on the other. Thus the spectra, which depend on the dust column 
density in the observable photosphere, 
depend on $T_{\rm eff}$ as well as on $T_{\rm cr}$. Then, the effects of 
$T_{\rm eff}$ and of $T_{\rm cr}$ cannot easily be separated on the observed 
spectra as well as on the infrared colors (Sect.2.1). Some examples of
such a degeneracy between  $T_{\rm eff}$ and  $T_{\rm cr}$ are given below.  

First, the observed spectrum of 2MASS\,1711 (L6.5) could be
fitted with the predicted one of $T_{\rm eff} = 1800$\,K on the assumption of 
$T_{\rm cr} = 1800$\,K (Fig.\,4a) as in Paper II. However, it is more 
likely that $T_{\rm cr} < 1700$\,K from the $J-K$ for this object (Table 1
and Fig.\,1b). Then, the same observed spectrum can be fitted with
the predicted one of $T_{\rm eff} = 1300$\,K  on the assumption of 
$T_{\rm cr} = 1700$\,K (Fig.\,4b). It is a bit surprising that a
difference of only 100\,K in $T_{\rm cr}$ results in a difference of 500\,K
in $T_{\rm eff}$. But another complication due to dust is coupled here: 
As can be known by the fact that the same infrared color corresponds to
two different values of $T_{\rm eff}$ (Fig.\,1), a very similar SED may
result from objects of different $T_{\rm eff}$ values, one relatively high
and the other relatively low, as discussed in Paper II for the case of
the L5 dwarf 2MASS\,1507 (see its Fig.\,13).
For this reason, not only $T_{\rm eff} = 1800$\,K but also a lower 
$T_{\rm eff}$ near 1300\,K could explain the overall SED for the case of
$T_{\rm cr} = 1800$\,K. The lower $T_{\rm eff} \approx 1300$\,K,
however, could be rejected because of the strong methane bands predicted for
this case
\footnote{Note that the predicted 
spectra show bifurcations in the region of methane bands near 1.6 and 
2.2\,$\mu$m according as the cases I (band model opacity) or II (linelist) 
opacities are used for CH$_4$. As for detail, see Paper II.}.
For $T_{\rm cr} = 1700$\,K, the situation is essentially the same
but the methane bands are not so strong even for the case of $T_{\rm eff}  
\approx 1300$\,K because of the stronger dust extinction due to the thicker 
cloud for the lower value of $T_{\rm cr}$, and there is no longer any
reason why to reject this case
\footnote{For the same reason, our previous interpretation in Paper II on
2MASS\,1507 (L5) should also be reconsidered, and the low temperature model 
with the lower value of  $T_{\rm cr}$ should apply rather than the high 
temperature model with $T_{\rm cr} = 1800$\,K.}.

As another example, the overall SED of SDSS\,1750 (T3.5) can be
fitted with the predicted one of $T_{\rm eff} = 1100$\,K on the assumption of 
$T_{\rm cr} = 1800$\,K (Fig.\,5a), although the predicted water bands 
appear to be too strong (Paper II). The infrared colors of this object 
(Table 1), however, suggest that $T_{\rm cr}$ should be quite high and may be 
as high as $T_{\rm cond}$ (see Fig.\,1b). Then, on the assumption of 
$T_{\rm cr} = T_{\rm cond}$ (case C or effectively no cloud), the same 
observed spectrum can be fitted with the predicted one of $T_{\rm eff}= 
1300$\,K and the water bands show even better agreements (Fig.\,5b).

\subsection{Reanalysis of the Spectral Energy Distributions}

The ambiguity due to the degeneracy of $T_{\rm eff}$ and $T_{\rm cr}$ in 
the analysis of the spectra cannot be removed if only their shapes, or 
relative SEDs, are 
analyzed as in Sect.3.1. This difficulty, however, may be relaxed to some 
extent if we reduce the observed spectra to the SEDs on an absolute scale.
With the known distance, $d$, by the recent astrometry measurements and with
the assumption that the radius, $R$, can be assumed to be the same with the
Jupiter's radius, observed SED, $f_{\nu}$(Jy), can be transformed to an 
absolute scale, i.e. the emergent flux from the unit surface area of the
object, $F_{\nu}$, by 
$$ {\rm log}\,F_{\nu}({\rm erg\,cm^{-2}\,sec^{-1}\,Hz^{-1}}) =
   {\rm log}\,f_{\nu}{\rm (Jy)} - 2 {\rm log}(R/d) -23.4971  . $$
Then, observed and predicted SEDs can be compared directly and 
no further ambiguity of vertical shift is left. We reanalyze the
spectra discussed in Paper II, but we reduce all the spectra to be 
analyzed in this subsection to the absolute scale with the use of the 
parallaxes by \citet[][]{vrb04} and assumption of $R = R_{\rm Jupiter}$. 
The largest source of uncertainty in this transformation is the assumption on
the radius, and a 30\% error in the assumed radius \citep{burr01} may result 
in an error of 0.2 - 0.3 dex on log\,$F_{\nu}$.

In Fig.\,6, the observed SED of the L6.5 dwarf 2MASS\,1711 represented
by the dots is compared with the predicted one based on the UCM with 
$T_{\rm cr} = 1800$\,K and $T_{\rm eff} = 1800\,K$ (labeled  A in the upper 
panel). This model was thought to fit reasonably well for this
object in Paper II (see its Fig.\,6). However, it now appears that the fit is 
very poor on the absolute scale, even though the overall shapes of the 
spectra agree rather well as shown in Fig.\,4a. For this
object, empirical value of $T_{\rm eff}$ is  1545\,K and $J-K = 2.25$
\citep{vrb04}. These data may suggest that $T_{\rm cr} \la 1700$\,K from 
Fig.\,1b. We then looked for a better fit in the SEDs predicted with the 
UCMs of $T_{\rm cr} = 1700$\,K and found that a model of $T_{\rm eff} =
1300\,K$  (B in the lower panel of Fig.\,6) shows a
relatively good fit, except that the water bands appear
to be too strong. Probably, this may not be a unique solution, but
this solution is more consistent with the available observed data.  

In Fig.\,7, the observed SED of the L8 dwarf 2MASS1523  
is compared first with the predicted one with the UCM of $T_{\rm cr} = 
1800$\,K and $T_{\rm eff} = 1500\,K$ (A in the upper panel). This model
was deemed as the best choice for this object based on the analysis of 
the same spectrum without absolute calibration in Paper II (see its Fig.\,8).
In the upper panel of Fig.\,7, the fit is very poor on the absolute scale, even
though the overall shape of the spectra agree rather well. For this object, 
empirical value of $T_{\rm eff}$ is  1330\,K and $J-K =1.65$ \citep{vrb04}.
These data may suggest that $T_{\rm cr} \approx 1700$\,K from Fig.\,1b.
We then looked for a better fit with the UCMs of $T_{\rm cr} = 1700$\,K
but found no reasonable solution. A model with $T_{\rm eff} = 1300\,K$, 
(B in the lower panel of Fig.\,7) shows a difference of as large as 0.2 dex on
log\,$F_{\nu}$ even though $T_{\rm eff}$ is close to the empirical value.
To have a better fit in the absolute scale, $T_{\rm eff}$ may be reduced 
further but it will make methane bands too strong and may not be acceptable.
Probably, the difference of 0.2\,dex may be within the uncertainty 
of the absolute flux due to our assumption on the radius as noted above, 
and we suggest that a possible solution can be $T_{\rm cr} \approx 1700$\,K 
and $T_{\rm eff} \approx 1300$\,K.

In Fig.\,8, we show the observed SED of T2 dwarf SDSS\,1254 by the dots 
and compared it with the predicted one based on the UCM of
$T_{\rm cr} = 1800$\,K and $T_{\rm eff} = 1300\,K$ (A in the upper 
panel). This is the model shown in Fig.\,9 of Paper II and suggested to 
be the best model based on the fit of the overall shapes of the observed
and predicted spectra. It is now confirmed that the fit is also fine in the 
absolute scale. Also, $J-K = 0.82$ \citep[][]{kna04} of this object may  
suggest that $T_{\rm cr} \approx 1800$\,K from Fig.\,1b.  The value of
$T_{\rm eff}$ is only slightly lower than the empirical
value of $T_{\rm eff} = 1361\,K$ \citep[][]{vrb04}. 
We examined the case of $T_{\rm eff} = 1400\,K$ which required
$T_{\rm cr} = 1900$\,K to reproduce the overall shape of the observed
spectrum. This case (B in the lower panel of Fig.\,8) shows
some deviation from the observed spectrum, although the deviation may be
within the uncertainty of the absolute scale as noted above.

Finally, we examine the case of T3.5 dwarf SDSS\,1750 in Fig.\,9, in which the
observed SED is compared with the predicted ones from the two different
UCMs. First, the case of $T_{\rm cr} = 1800$\,K and $T_{\rm eff} = 1100\,K$,
which we have accepted in Paper II, reproduces the relative SED reasonably 
well except that water bands appear to be too strong, and we have further
suggested log\,$g$ = 5.5 to reduce the water band strengths in Paper II 
(see its Fig.\,10). The fit in the absolute scale for this model is not so good
in Fig.\,9 even though it may be within the uncertainty of the absolute flux
(A in the upper panel). For comparison,
recent empirical value is $T_{\rm eff} = 1478\,K$ \citep{vrb04}, about 400\,K
higher than the value based on the model fitting. Also, this object is as 
blue as $J-K = 0.12$ \citep{kna04} and the dust cloud may have cleared already
at T3.5 as can be inferred from Fig.\,1b. We then apply our model with
$T_{\rm cr} = T_{\rm cond}$ (case C) and looked for the best fit. We found 
that a model of $T_{\rm eff} = 1300\,K$ provides the best fit  both in
relative and absolute scales (B in the lower panel of Fig.\,9).
Another photometry suggests that $J-K = 0.83$ \citep{vrb04}
and we also examined the case of $T_{\rm cr} = 1900$\,K. It turns out
that the lower values of $T_{\rm cr} $ and/or the higher values of 
$T_{\rm eff}$ do not improve the fit in relative as well as in absolute scale.

The possible best values of $T_{\rm eff} $ based on the reanalysis outlined
above are summarized in Table 1 together with some observed data, and
will be discussed in Sect.3.3. 

\subsection{Interpretation of the L-T Spectral Sequence}

As an example of the L-T spectral sequence, we reproduce in Fig.\,10 
the spectra of cool dwarfs from L6.5 to T3.5, which were discussed
in some detail elsewhere (Nakajima, Tsuji, \& Yanagisawa 2004).
The spectra show systematic changes from the L dwarfs with H$_2$O
bands of modest strengths and marginal CH$_4$ bands to
the T dwarfs with strong bands of both H$_2$O and CH$_4$. Also, the
overall SEDs in the spectral range of Fig.\,10 change from red in the L 
dwarfs to blue in the T dwarfs. 
This spectral sequence could have been interpreted as a temperature
sequence which extends from 1800 to 1100K ( reproduced in the
6-th column of Table 1), so long as we assume a uniform value of 
$T_{\rm cr} = 1800$\,K throughout L and T dwarfs (Paper II).
This seemed to be a reasonable result for such a large and systematic 
change of the spectra, and also fitted to the general expectation
that the L-T spectral classification may be an extension of
the stellar spectral types beyond M to the cooler temperatures.

In Sect.3.2, we reanalyzed the
same data reproduced in Fig.\,10 with the $T_{\rm cr} $ values inferred from 
the $J-K$ values of individual objects (Table 1 and Fig.\,1b). Also fitting of
the observed and predicted spectra is done not only by the relative
shape of the spectra but also by the use of the absolute flux scale
based on the recent astrometry data.  The result of 
the reanalysis turns out to be quite different as summarized in the 7-th
column of Table 1. The spectra from L6.5 to T3.5 shown in Fig.\,10 can be 
interpreted by a uniform value of $T_{\rm eff} \approx 1300$\,K throughout. 
Instead, the spectral sequence of Fig.\,10 can be explained by the 
change of $T_{\rm cr}$ from about 1700\,K to $T_{\rm cond}$ (i.e. case C) 
\footnote{In this interpretation, L6.5 dwarf 2MASS\,1711 and L8 dwarf 
2MASS\,1523 have the same $T_{\rm eff}$  (Figs.\,6 
\& 7), but we are now discussing the first order effect with models 
in which $T_{\rm eff}$ and $T_{\rm cr}$ are changed by a step of 100\,K,
and a further fine tuning should certainly be needed.}.  An important
conclusion is  that the effect of $T_{\rm cr}$ is sometimes more important 
than that of $T_{\rm eff}$, and such drastic change of spectra that requires
different spectral types L and T can be due to the change of $T_{\rm cr}$ 
rather than of $T_{\rm eff}$.

Thus, the present reanalysis shows a marked contrast to the previous 
analysis (Paper II), and it is quite unexpected  that the L-T spectral
sequence can no longer be regarded as a temperature sequence, but
represents a sequence of the change of the dust column density in the
observable photosphere at least between the middle L and early T dwarfs.
Nevertheless, we should accept the new result which is based on the
analysis of the SEDs on an absolute scale. 
Even though the absolute flux may be uncertain by as much as  
$\Delta$log\,$F_{\nu} \approx 0.2 - 0.3$\,dex because 
of the uncertainty of the radius of an individual object as noted in Sect.3.2,
the ambiguity in the analysis of the relative SED should be still larger
as shown in Sect.3.2.  Also, the critical temperature  $T_{\rm cr}$  is 
already selected to be consistent with the infrared colors (Fig.\,1b), and  
the resulting $T_{\rm eff}$ values
agree with the recent empirical effective temperatures which suggested 
an almost uniform value of $T_{\rm eff} \approx 1400$\,K in the same spectral
range \citep[][]{vrb04, gol04}. A systematic difference of 100K remains to be 
explained, but such a difference may be within the error bars of the both 
analyses.

\section{BASIC PHYSICAL PROPERTIES OF ULTRACOOL DWARFS}
Recent progress in observations of such faint objects as L and T dwarfs
is quite substantial, especially in photometry \citep[e.g.][]{leg02, gol04,
kna04} and astrometry (e.g. Dahn et al. 2002; Tinney, Burgasser \& Kirkpatrick
2003; Vrba et al. 2004). Thus, we now have reasonably accurate  empirical
data on luminosities and effective temperatures for a large sample of L and T
dwarfs. At the same time, these new data uncovered a difficulty in
the spectral classification of L and T dwarfs, which was
essentially following the methodology of the stellar spectral classification.

\subsection{Luminosities}
 Recent progress in observations finally provided reasonably accurate
absolute magnitudes (monochromatic as well as bolometric) for a large
sample of L and T dwarfs \citep[e.g.][]{dah02, tin03, vrb04}. As an example, we
reproduce the  $M_{J}$ plotted against the  L-T spectral type in Fig.\,11a
with the data by \citet[][]{vrb04} (see their original plot as for more 
detail with error bars),
showing the so-called $J$ -brightening at the early T dwarfs.
It is known that the situation is more or less the same for $M_{H}$
and $M_{K}$ if not so pronounced as in $M_{J}$  
\citep[e.g.][]{dah02, tin03, vrb04}. Then, these data may suggest
that the bolometric magnitude $M_{\rm bol}$ should also show the similar
brightening, since most of the flux is emitted in $J, H,$ and $K$ bands.
We confirmed in Fig.\,11b that the absolute bolometric magnitude 
$M_{\rm bol}$ plotted against the L-T spectral type with the data from 
\citet[][]{vrb04} in fact shows the brightening at the early T dwarfs.   
The values of bolometric magnitude $M_{\rm bol}$ used in Fig.\,11b are already 
based on the bolometric corrections by \citet[][]{gol04} including the 
result of the $L^{'}$ and $M^{'}$ photometry. However, we examine the
$M_{\rm bol}$ values of \citet[][]{gol04} based on a different photometric 
system in Fig.\,11c. The ``brightening'' may not be so clear as in Fig.\,11b,
but the $M_{\rm bol}$ also levels off between L5 and T4 in Fig.\,11c.

Although these results are quite unexpected, it is more easy to
interpret Fig.\,11b (or Fig.\,11c) than Fig.\,11a:
Since the bolometric luminosities essentially reflect the cooling of
brown dwarfs, the brightening (or the level-off) of the total luminosities is 
difficult to understand if they are plotted against a correct temperature 
indicator. An only possible solution may be that the problem is not in the
ordinate but in the abscissa of Fig.\,11b (or Fig.\,11c), namely  the L-T 
spectral sequence is not representing a temperature sequence.
Since Fig.\,11a and Fig.\,11b are based on the same photometry data and show
quite similar ``brightening'', the brightening in Fig.\,11a should most 
probably be due to the same origin as in Fig.\,11b. This result implies that 
the so-called ``$J$ - brightening'' is not due to any atmospheric effect,
but should be due to intrinsic properties of the early T dwarfs. 
This result that the L-T spectral sequence is not a temperature sequence 
is consistent with the recent empirical effective temperatures 
\citep[e.g.][]{gol04, vrb04} and with the conclusion of Sect.3.3 based
on the analysis of a limited sample of L and T dwarfs.
All these results in turn lend a support to
our interpretation on the ``brightening'' . Thus, the so-called 
``$J$-brightening'' problem on the $M_{J}$ - Sp.\,Type plot no longer exists, 
while that on the CM diagram has been solved already (Sect.2.3).

\subsection{Effective Temperatures}

Effective temperature is one of the basic fundamental parameters
that specify the 
nature of substellar as well as of stellar object. Accurate determination
of $T_{\rm eff}$, however, is generally difficult because the total energy 
flux integrated over the whole spectral region is required on an absolute
scale (i.e. in unit of energy flux emerging from the unit surface area
of the object). In the case of star, observed bolometric flux can be
converted to the absolute emergent total flux if the stellar
angular diameter could be measured. In the case of brown dwarf, the angular 
diameter is more difficult to measure at present but its
equivalence can be estimated if the distance can be known, because of the
favorable nature of brown dwarfs that the radius is within 30\%
of the Jupiter's radius independently of mass \citep[e.g.][]{burr01}.  

Recently, reasonably accurate trigonometric parallaxes are
made available thanks to the elaborate astrometric observations 
by several groups \citep[e.g.][]{dah02, tin03, vrb04}. 
These results are combined with the photometric observations
well extended to the $L'$ and $M'$ regions \citep[e.g.][]{leg02, gol04, kna04}
and empirical effective temperatures as well as the absolute magnitudes
are determined for dozens of L and T dwarfs. An important finding is that the 
empirical $T_{\rm eff}$'s by \citet[][]{vrb04} and by \citet[][]{gol04} 
show a large plateau between
about L5 and T5 and, especially, $T_{\rm eff}$ between L8 and T2 is nearly 
constant. This result was also suggested by \citet[][]{nak04}
based on the bolometric correction predicted with the use of the UCMs. 
The effect of log\,$g$ may not be so large as
to modify the qualitative nature of the spectra.
Thus the L/T transition as well as the change of the spectra between late L 
and early T types should not be due to $T_{\rm eff}$ and log\,$g$ alone but 
may be governed by another parameter, which we have already identified
with $T_{\rm cr}$.

With the empirical $T_{\rm eff}$ as the abscissa, we  plot  $M_{J}$ 
in Fig.\,12a. Since $T_{\rm eff}$ is derived from $M_{\rm bol}$ which is 
not fully independent of $M_{J}$, it is expected from the beginning 
that  $M_{J}$ and $T_{\rm eff}$ are somewhat related. Nevertheless,
if the $J$-band flux suffers a serious atmospheric effect, such a smooth
monotonic curve as Fig.\,12a may not necessarily be trivial. It is confirmed
that the ``brightening'' such as shown in Fig.\,11a disappears if $M_{J}$ is 
plotted against effective temperature, and this fact can be regarded as a 
confirmation of the conclusion of Sect.4.1 that the so-called
``$J$-brightening'' is  a problem of L-T classification rather than of
$M_{J}$. Also, it is to be noted that the scattering is rather small 
in Fig.\,12a except for the L/T transition region. 
We overlaid in Fig.\,12a the evolutionary tracks of 10, 42, and 
70\,$M_{\rm Jupiter}$  transformed to $M_{J} - T_{\rm eff}$ from $M_{\rm bol}
- T_{\rm eff}$ \citep[][]{burr97} for different values of $T_{\rm cr}$. 
It can be confirmed that the variable $T_{\rm cr}$ in fact results 
in the scatters in $M_{J}$  at $T_{\rm eff} \approx 1500$K where the effect 
of the dust clouds is the largest, but that  the ``atmospheric effect'' is 
not so large as to produce the so-called ``$J$-brightening''.
 
 To investigate the nature of the empirical $T_{\rm eff}$, we also plot
$M_{\rm bol}$ against the empirical $T_{\rm eff}$ (error bars in the original 
data are not shown, but typically $\pm 200$\,K) in Fig.\,12b, and
the evolutionary tracks for 10, 42, and 70 $M_{\rm Jupiter}$ \citep{burr97}
are overlaid.   The empirical values of $T_{\rm eff}$ are derived with the 
assumption of the uniform radius of  0.90$R_{\rm Jupiter}$ by 
\citet[][]{vrb04} with their eqn.(7), and the results seem to be 
approximately equivalent to have derived $T_{\rm eff}$ with 
the use of the evolutionary tracks for 42 - 70\,$M_{\rm Jupiter}$. Thus, the 
empirical $T_{\rm eff}$'s are  subject to some ambiguities due to the 
uncertainties of mass, radius, and/or age. Other authors \citep[e.g.][]{
gol04, kna04} applied the evolutionary tracks to convert $M_{\rm bol}$ to 
$T_{\rm eff}$, but the exact evolutionary status
of an individual object is anyhow unknown. For example, \citet[][]{
gol04} showed that the range of possible $T_{\rm eff}$ derived from 
$L_{\rm bol}$ assuming evolutionary models with ages of 0.1 - 10\,Gyr 
is $\approx 300$\,K. In Fig.\,12c, we also plot $M_{\rm bol}$ against the 
empirical $T_{\rm eff}$ derived for an age of $\approx$\,3Gyr 
by \citet[][]{gol04}. Except for a few cases, Fig.\,12b and
Fig.\,12c appear to be rather similar, and it looks as if most brown dwarfs
are on the evolutionary tracks of 42-70\,$M_{\rm Jupiter}$. However, this 
is simply a consequence that the age of most brown dwarfs are assumed to be 
$\approx$\,3Gyr. 

More or less similar difficulty has been noticed by \citet{gol04}, who 
compared the diagrams of $M_{K}$, $M_{L'}$ and  $M_{M'}$  
versus $T_{\rm eff}$ with the predictions based on the
models of \citet{mar02} for varying sedimentation 
efficiencies $f_{\rm sed}$ (3, 5, and no cloud) and different values of
log\,$g$ (4.5, 5.0, and 5.5). 
Their results showed that the model predictions reproduce well
the broad ranges of observed absolute magnitudes and effective temperatures
for a large sample of L and T dwarfs. Yet it appeared to be difficulty to 
decide particular sets of model parameters because of the difficulty of
knowing masses, ages, and metallicities of individual objects.
\citet{gol04} also showed that the L3 - T4.5 dwarfs generally have higher
gravities than the T6 - T9 dwarfs, and these results may be consistent with  
our  Fig.\,12 in that earlier dwarfs generally tend to populate the 
evolutionary tracks of the higher mass models. 

Despite the difficulty of observing such faint objects as brown dwarfs,
one relief is that parallaxes could have been measured accurately.
In fact,  the recent achievement of measuring the
parallaxes of dozens of ultracool dwarfs is the most important contribution
to our understanding of brown dwarfs (e.g. see Sect.2.1 and Sect.4.1).
Once distance of object can be known accurately, it could be expected
that we can understand well the nature of astronomical object. 
Nevertheless, it appeared that the empirical $T_{\rm eff}$  
values for individual objects are still subject to uncertainty of
about $\pm 200$\,K \citep{vrb04,gol04,kna04} and, for this reason, 
it is still difficult to determine masses and/or ages of individual
objects from the observed data at present.
We must wait for direct measurements of radii or angular
diameters for a final solution of this uncertainty. As a 
compromise, model analyses such as given in Sect.3 can be of some use
to examine the consistency of various observed data, and to improve   
our knowledge on $T_{\rm eff}$ and other parameters.

\subsection{Spectral Classification}

A conclusion to be drawn from Sects\, 3.3 and 4.1 is 
that the L-T spectral classification may have serious problem regarding
its meaning as a temperature classification. The stellar spectral 
classification has been done on purely empirical basis and
it was not  intended to be a temperature sequence from the
beginning.  However, the great success of the stellar spectral classification
established at the beginning of the 20-th century is largely due to the
finding that the spectral sequence is a temperature sequence.  Later efforts
to interpret the stellar spectral sequence in terms of temperature and 
other physical parameters finally resulted in establishing the present-day 
stellar astrophysics. The present L - T spectral classification has also 
been done  on purely empirical basis \citep[e.g.][]{kir99}
and again it was not explicitly
mentioned that the resulting spectral sequence will be
 a temperature sequence.
Thus it is left free as to how to interpret the L - T spectral types.
However, the L and T types are conceived as extensions of the spectral type
 beyond M, and may implicitly be expected to be temperature indicators. 

Anyhow, we should first understand what the present L - T spectral 
types mean.
Although the reason for the difficulty to interpret the L - T spectral types
can be explained as due to the degeneracy of $T_{\rm eff}$ and $T_{\rm cr}$ 
on the spectra contaminated with dust (Sect.3.1), at least partly, 
this problem may be most serious in the region around 
$T_{\rm eff} \approx 1400 \pm 100$\, where the transition from L to T
takes place (Fig.\,1). In other regions, however, the situation may be
different, and the L - T spectral classification may be used as a temperature
classification  if applied with some cautions. We now apply our UCM 
as a working model to interpret the L-T classification in three different
parts, namely, early L dwarfs (roughly $T_{\rm eff} > 1500$\,K),  L-T
transition region (1300\,K $\la T_{\rm eff} \la 1500$\,K), and
dust-cleared T dwarfs ($T_{\rm eff} < 1300$\,K):

In the early  L dwarfs, the dust clouds that occupy the region of
$T_{\rm cr} \la T \la  T_{\rm cond}$  are entirely located in the optically 
thin region.
In this case, the location of the lower boundary of the cloud at $T \approx 
T_{\rm cond}$ should be more important than that of the upper boundary  
at $T \approx T_{\rm cr}$, since the region near the upper boundary of the 
dust cloud will have little contribution to the dust column density
because of the lower density there. For this reason, the spectra essentially
 depend on $T_{\rm cond}$ which in turn can be well interpreted in terms of 
$T_{\rm eff}$ and log\,$g$, and this explains why scatters are small in
the early L dwarfs in Fig.\,1.  At the same time, it is difficult to known 
the value of $T_{\rm cr}$ from the infrared colors (Fig.\,1b).

From middle L to  early T dwarfs, the L-T spectral
classification met serious difficulty as noted above. In this region, 
the lower boundary of the cloud at $T \approx  T_{\rm cond}$ will be in
the unobservable photosphere (i.e. in $\tau > 1$), and the thickness of
the cloud in the observable photosphere is essentially determined by 
$T_{\rm cr}$, which is not controlled by $T_{\rm eff}$ nor by log\,$g$. 
For this reason, the spectra show little effect of $T_{\rm eff}$ 
but depend mostly on $T_{\rm cr}$, and an actual example of this case
is shown by the spectral sequence reproduced in Fig.\,10. The analysis outlined  
in Sect.3.2 is based on a limited sample which, however, may represent
the general feature of the spectra between middle M and early T types.
This is because a larger sample shows the same conclusion: First, empirical 
$T_{\rm eff}$ plotted against spectral types shows a large plateau in the 
middle part of the L-T sequence \citep{vrb04, gol04, nak04, leg05}. Second, 
$M_{\rm bol}$ plotted  against spectral types shows a ``brightening'' around 
early T types (Fig.\,12b), which is difficult to understand if the spectral 
types are on a correct temperature sequence. Thus the spectral classification 
between middle L and early T types may radically be reconsidered. 

In the middle and late T dwarfs,  the effect of cloud is diminishing because 
of the immersion of the cloud in the optically thick region, and gaseous
opacities dominate. The upper boundary of the cloud may still in the 
observable photosphere in the middle T, but it  seems that $T_{\rm cr}$ may 
be systematically high in these dwarfs (about 1900K or higher from Fig.\,1b 
and Fig.\,2b). This fact may suggest a possibility that the dust cloud is 
already quite thin in middle T dwarfs, but the reason for this is unknown.
Finally, in the late T dwarfs, dust cloud is in the optically thick 
region and  no information on the cloud can be
obtained from observations \citep[][]{lie00}. In this case, 
$T_{\rm cond}$ as well as $T_{\rm cr}$ has no effect on spectra as well as
on colors. The scatters in colors at the fixed
$T_{\rm eff}$ are mostly due to gravity effect as noted in Sect.2.

Apart from the problems related to $T_{\rm cr}$,
one difficulty in the classification of dusty dwarfs is that
the dust itself shows no clear spectral signature and it is very
difficult to estimate the dust column density directly from observations.
The spectral classification had to be done with the use of the
spectral features originating from  atoms and molecules, while
the spectra are largely controlled by dust. For example, methane bands
which are used in the infrared classification  show drastic change from
L to T dwarfs (e.g. Fig.\,10) but this is not due to a direct effect of
the change of the methane abundance with $T_{\rm eff}$ but rather due to an 
indirect effect of the change of the dust extinction with $T_{\rm cr}$ as
discussed in Sect.3.3. 
But there should be a case that the methane bands actually change
due to the change of the methane abundance with $T_{\rm eff}$.
For this reason, methane bands cannot be indicators of temperature, and the 
situation may be more or less the same for other atomic and molecular features.

Another difficulty of the present L - T classification has also been
pointed out in that the L types based on the optical spectra and those
on the near infrared spectra show differences as large as 3 subtypes
\citep[e.g.][]{ste03, kna04, leg05}. This fact, however, may offer an 
interesting possibility of the two-dimensional spectral classification
of ultracool dwarfs, if the optical types can be indicators of temperature 
and the infrared types of the dust opacities \citep[][]{ste03}. In view of the 
difficulties outlined above, however, it seems to be more difficult to
separate the effects of temperature and of dust opacities.
For example, the effect of the dust column density is more serious as to
produce such a large difference in spectral types from L6.5 to T3.5
for about the same  $T_{\rm eff}$ as shown in Fig.\,10 rather than results
in just 3 subclasses at the largest in late L dwarfs.

The present L-T classification is essentially following the methodology 
of the stellar spectral classification \citep[e.g.][]{kir99}, which is
a marvelous  art to infer  the basic stellar properties 
such as the effective temperature and luminosity 
by just looking at the low resolution spectra.
In the case of the spectra of cool dwarfs  contaminated with dust, such a 
convention may be more difficult not only because an additional parameter
related to the dust clouds is required to characterize the spectra
 but also the dust itself shows no spectral signature.
As a parameter to specify the property of the dust cloud, we introduced
$T_{\rm cr}$, which is a measure of the dust column density in our UCM,
but the real difficulty is that the effects of $T_{\rm eff}$ and $T_{\rm cr}$ 
cannot be separated even if the effect of log\,$g$ could be known. 
It is of course desirable to be able to know the basic
parameters such as  $T_{\rm eff}$, log\,$g$,  and metallicity, and
further some dust properties (e.g. $T_{\rm cr}$ in our UCMs) by
just looking at the low resolution spectra. However, it seems to be a 
formidable problem to have a  classification scheme similar to
the stellar spectral classification for the spectra of dusty dwarfs, and
more detailed and careful examination of the spectra from optical to 
infrared region should be required.

\section{DISCUSSION AND CONCLUDING REMARKS}

Recent observations of ultracool dwarfs revealed some confusion
as to their interpretation. For example,
the ``brightening'' in the infrared absolute magnitudes 
at the early T dwarfs on the L-T sequence as well as on the  CM diagrams 
\citep[e.g.][]{dah02, tin03, vrb04}, large scatters in the infrared colors 
(Knapp et al, 2004 and references cited therein), a large plateau 
in $T_{\rm eff}$ - Sp.\,Type calibration  between middle L and middle 
T \citep[][]{vrb04} etc.  We have shown in this paper that these difficulties 
are mostly resolved if we apply our UCMs properly, i.e., if we allow
the critical temperature to change from as high as the condensation 
temperature to near the surface temperature.  We have not yet considered 
another curious observation that the absorption bands of FeH decrease first 
from early to late L types but strengthen again at early T types 
\citep[][]{bur02b, nak04}. We have no enough data to analyze this 
problem in detail, but we suggest that this may be the same phenomenon as the
``$J$-brightening'' at the early T types. In other words, if the FeH
band strength is plotted against $T_{\rm eff}$ rather than L-T types,
it will show a steady decrease and we hope that this will be confirmed
by the future observations. Also, if this is confirmed, FeH can be a useful
marker for spectral classification.   

Within the framework of UCM,  the critical temperature  $T_{\rm cr}$ shows 
such a large and unpredictable variation at a given $T_{\rm eff}$, and this 
result implies that the dust column density in the observable photosphere 
differs largely for the same $T_{\rm eff}$ (and other basic parameters).
It is remarkable that
the variation of $T_{\rm cr}$ is especially large at $T_{\rm eff} \approx
1400 \pm 100$\,K, and T dwarfs in this region mostly show $T_{\rm cr}
\ga 1800$\,K while L dwarfs $T_{\rm cr} < 1800$\,K (Fig.\,1b).
Then  $T_{\rm cr}$ is as high as 1900\,K or near $T_{\rm cond}$ 
in middle T dwarfs as noted in Sects.2.2 and 4.3. These results may
indicate that the thickness of the dust cloud itself is decreasing after
the L-T transition and further throughout T dwarfs. Thus it appears
that thinning of the dust cloud should be associated with
the immersion of the cloud in T dwarfs, but it is not clear if this
is also the case  in  L dwarfs.   A problem is if the dust cloud
will finally disappear in late T dwarfs or if it still exists in the deeper 
layer. Unfortunately, this problem cannot be answered with observations,
since no information is available from the optically thick region.

It is to be remembered that the immersion of the dust cloud in cooler
dwarf is a natural consequence of the change of the photospheric structure
according as L dwarf evolve to T dwarf, and this is simply because
the dust cloud always forms at about the same temperature, namely,
at $T_{\rm cond} \approx 2000$\,K, where is in the optically thin (thick) 
region of the objects with relatively high (low) $T_{\rm eff}$. However, the 
thinning of the dust cloud is more difficult to understand by such a simple 
picture. A known noticeable change of the photospheric structure at 
$T_{\rm eff} \approx 1400$\,K is the formation of the second (outer) 
convective zone (Paper I), which may have some impact on the formation
and destruction of the clouds.  Since the convective activities 
may depend on the age and evolutionary history of the cooling brown dwarfs,
the nature of cloud including $T_{\rm cr}$ may depend on such effects as well. 
However, the details of how the convective activity is related to 
the variation of $T_{\rm cr}$ is unknown at present.

The critical temperature $T_{\rm cr}$ and its variation should be a key in our 
understanding of the photospheric structure and hence of the observed 
properties of dusty dwarfs. In this paper, however, we leave $T_{\rm cr}$  
as a free parameter to be estimated from observations.
It is of course desirable that $T_{\rm cr}$, or
more generally the upper boundary of the dust cloud, can be determined
from the basic physics. However, this will anyhow be highly model dependent
at present. For example, \citet{woi04}  recently proposed an interesting 
model in which a series of processes such as  nucleation, dust growth, 
gravitational  settling, evaporation, and element replenishment will take 
place in a convective life cycle in the photosphere, and the structure of 
the dust cloud including its upper boundary as well as dust properties
such as the grain size can be determined by solving their moment equations
in the circulating flow. In this model, however, everything depends on  
a parameter referred to as the mixing timescale, which was assumed as a
measure of the efficiency of the convective activity including overshooting.
But it is by no mean clear if the mixing will take place  in such a way
as to interact with the dust formation so nicely as assumed  in their model. 
For example, the structural models of cool dwarfs show that convective zone 
is situated deeper in L dwarfs than in T dwarfs \citep[e.g.][]{tsu02},
and hence mixing in the photosphere induced by the convection  will be 
more effective  in T dwarfs than in L dwarfs. Then, if the mixing  plays a 
major role in cloud formation, a question is why T dwarfs are not dusty 
while L dwarfs are.

Our major concern at present is to interpret observed data in terms of the 
basic stellar parameters such as $T_{\rm eff}$, log\,$g$, metallicity, 
chemical composition etc.
For this purpose, it is required to predict
observables such as colors, magnitudes, spectra etc. and our 
empirical model referred to as UCM is primarily developed for such 
applications in mind. So far as the model is empirical, the cloud property
such as $T_{\rm cr}$ (i.e. the upper boundary of the cloud) is simply 
estimated empirically and hence free from any particular model
of cloud formation, which is presently not well established yet.
Such an empirical approach still plays an important role in our studies
of stellar and substellar photospheres.  For example, 
stellar convection theories, which we have also employed in our UCMs,
mostly involve a parameter known as the mixing 
length, and the micro-turbulent velocity is usually determined empirically 
rather than evaluated theoretically for individual object. 
Of course an empirical approach is only an initial step toward more complete
modeling, but we must recognize that the stellar and substellar photospheres 
still involve formidable problems for which fully theoretical solution is 
difficult at present.

Finally, it is to be noted that the critical temperature $T_{\rm cr}$ is
a parameter inherent to our UCMs, and the fifth parameter next to  
the $T_{\rm eff}$, log\,$g$, chemical composition, and micro-turbulent 
velocity may not necessarily be restricted to $T_{\rm cr}$.
For example, other parameter that characterizes the dust clouds,
such as the sedimentation efficiency $f_{\rm sed}$ in the models of 
\citet[][]{mar02}, will play similar role as our $T_{\rm cr}$, and their 
cloudy models have extensively been applied to the interpretation and 
analyses of the recent observations of L and T dwarfs 
\citep[e.g.][]{kna04, gol04}.  From these analyses and from ours outlined in
this paper, the cloudy models of \citet{mar02} and our UCMs basically  
agree in that the cloud properties depend on 
a dynamical process whose origin cannot be identified yet and whose effect
had to be represented by a variable parameter such as $f_{\rm sed}$ or
 $T_{\rm cr}$.  In practical applications,
our $T_{\rm cr}$ may have some advantages in that it can easily be
incorporated in the classical non-grey theory.  Also, $T_{\rm cr}$ has a clear
physical meaning related to the thickness of the dust clouds, and hence
can be inferred directly from the observed infrared colors. Furthermore, 
we have shown that almost all the data points in the large collective
data such as the two-color diagram and CM diagram
can be consistently interpreted with the $T_{\rm cr}$ values in the
appropriate range, and this fact implies that $T_{\rm cr}$ is a
reasonable choice as a parameter for the characterization of dusty
dwarfs. 

In conclusion, we believe that we have now applied our UCMs 
more properly to the interpretation and analysis of the observed data 
of dusty dwarfs, with the recognition that the  parameter
$T_{\rm cr}$ takes different values in addition to $T_{\rm eff}$ and 
log\,$g$ (under the fixed chemical composition and micro-turbulent velocity). 
We certainly realized that dust plays an important role in
determining the observed properties of cool dwarfs at an early time,
but we did not fully understand its effect nor treated it properly
until now.  We should now recognize that the dust is one of the fundamental 
elements in the characterization of dusty dwarfs. Based on this
recognition, we must reexamine the observed data in more detail and 
reconsider such a basic problem as the
spectral classification of ultracool dwarfs, which cannot be a simple
extension of the stellar spectral classification. 
A relatively short history of the studies on ultracool dwarfs including
brown dwarfs was already quite intriguing, and thus such ultracool dwarfs will
remain to be  exciting subjects of further observational and theoretical
challenges.  

\acknowledgements
I thank Tadashi Nakajima for useful discussion and helpful comments on the
draft of this paper, and Mark Marley (as the referee) for constructive 
criticism and invaluable suggestions that helped to improve the initial 
version of this paper. Data analysis were in part carried out on general 
common use computer system at the Astromonical Data Analysis Center, ADAD,
of the National Astronomical Observatory of Japan.  This work was supported 
by the ADAC and by Grant-in-Aid 15204010 of JSPS (PI: M. Fujimoto).

\newpage
{\bf {\it Note added in proofs:}}
One difficulty in the present computation of the spectra of ultracool
dwarfs was the lack of reliable linelist of methane, but we were able 
to access 
a new extensive linelist by R. Freedman (private communication) after 
this work was completed. This linelist by Freedman has been generated
based on the software package "The Simulation of XY$_4$ Spherical Top 
Spectra" developed at Laboratoire de Physique de l'Universit\'e de 
Bourgogne\footnote{http://www2.u-bourgogne.fr/LPUB/TSM/sTDS.html}. 
We found that the new linelist (consists of about 10 million lines) 
offered substantial improvement over the previous linelist as in 
the GEISA database (consists of only some 50 thousand methane lines), 
and our case II spectra based the GEISA linelist should now be replaced 
with the ones based on the Freedman's linelist (to be referred to as 
case III). The spectrum computed with the case III methane opacity 
just comes between those computed with the case I (representing the 
maximum estimate with the completely smeared out band model) and the case 
II (under-estimate with only low excited lines) opacities as expected. 
This fact shows that the Freedman's linelist is an important step toward 
better methane opacity. However, we found that the methane opacity in the 
$H$-band region may be still underestimated with the Freedman's linelist 
and this fact may imply the extreme complexity of the methane spectra 
especially in the combination bands such as found in the $H$-band region. 
Instead the spectra in the $H$-band region could still be better represented 
by our case I spectra based on the band model opacity for methane. In fact, 
such colors as $J-H$ and $H-K$ could be reasonably well reproduced with 
the case I methane opacity as already shown in Fig.1. Thus, we suggest that 
the spectra and colors based on the Freedman's line list and/or those based 
on the band model opacity should be considered case by case in actual 
applications. For this purpose, we uploaded the predicted spectra and colors 
based on the case I and case III methane opacities in our website recently 
updated\footnote{http://www.mtk.ioa.s.u-tokyo.ac.jp/\~\,ttsuji/export/ucm
(updated on 05 Feb. 2005)}. 

   The update of the predicted spectra and colors owes very much to the 
generous help of Richard Freedman and Mark Marley, and I also thank 
them for invaluable discussion on the methane opacity.

\clearpage

\begin{table}
\caption{ EFFECTIVE TEMPERATURES}
\begin{tabular}{llccclll}
\noalign{\bigskip}
\tableline\tableline
\noalign{\bigskip}
Object & Sp. type & $J-K$\tablenotemark{a}  & $J-K$ 
\tablenotemark{b} & $d$ (pc)\tablenotemark{c}  &~ ${ T_{\rm eff} 
(T_{\rm cr})}$\tablenotemark{d}~ &
  ~ ${ T_{\rm eff} (T_{\rm cr})} $\tablenotemark{e}~ &  $ T_{\rm eff} $ 
\tablenotemark{f} ~ \\
\noalign{\bigskip}
\tableline
\noalign{\bigskip}
 2MASS\,1711+22 & L6.5 & -  & 2.25 & 30.20 & 1800\,K (1800\,K) & 1300\,K 
(1700\,K) & 1545\,K\\
2MASS\,1523+30 & L8 & 1.60 & 1.65 &17.45  & 1500~~~(1800) & 1300~~~(1700) &
 1330~~~\\ 
SDSS\,1254-01 & T2 & 0.82 & 0.96 & 13.21 & 1300~~~(1800) & 1300~~~(1800)  & 
1361~~~ \\
SDSS\,1750+17 & T3.5 & 0.12 & 0.83 & 27.59 & 1100~~~(1800) & 1300~~~(${ T_{\rm cond}}$) & 1478~~~   \\
\noalign{\bigskip}
\tableline
\end{tabular}
\tablenotetext{a}{MKO system (Knapp et al. 2004)}
\tablenotetext{b}{CIT system (Vrba et al. 2004)}
\tablenotetext{c}{distance based on the parallaxes by Vrba et al.(2004)}
\tablenotetext{d}{based on the analysis of the spectra with UCMs of the 
uniform value $T_{\rm cr} = 1800$\,K (Paper II)}
\tablenotetext{e}{based on the analysis of the absolute fluxes with UCMs of 
$T_{\rm cr} $ inferred from the infrared colors (this paper)}
\tablenotetext{f}{based on the bolometric luminosity (Vrba et al. 2004)}
\end{table}

\begin{figure}
\epsscale{1.00}
\vspace{-5mm}
\plotone{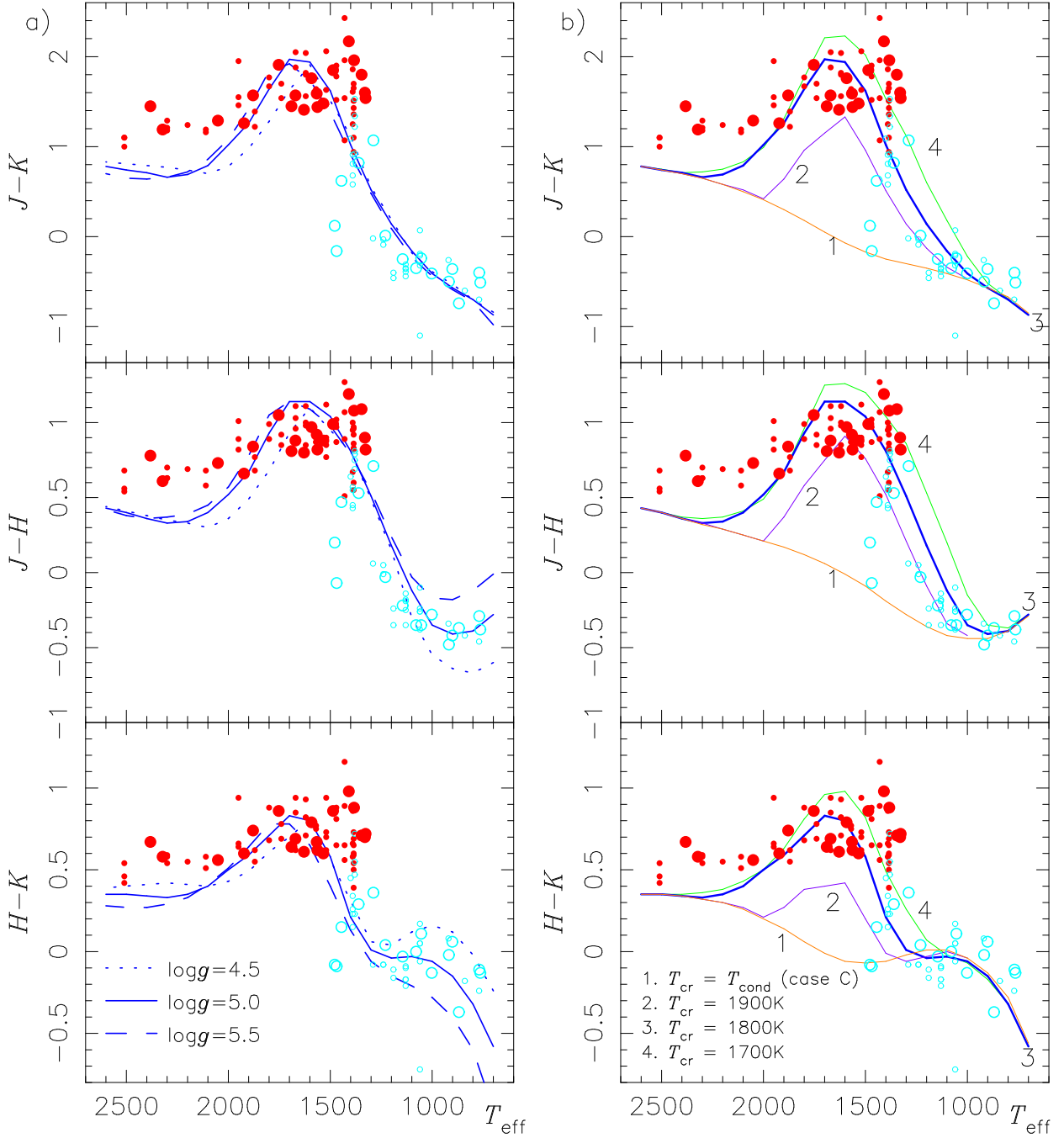}
\vspace{-5mm}
\caption {
Infrared colors on the MKO system (Knapp et al. 2004) are plotted against  
the effective temperatures (Vrba et al. 2004) by the filled and open circles 
for L and T dwarfs, respectively.
The large and small circles indicates that $T_{\rm eff}$ values are
determined directly from the measured bolometric luminosities 
and indirectly by the use of $T_{\rm eff}$ - Sp. Type calibration
by Vrba et al.(2004), respectively. $J-K, J-H,$ and $ H-K$ are shown
in the upper, middle, and lower panels, respectively.
a) Predicted colors based on the UCMs with log $g$ = 4.5, 5.0 and 5.5 are 
shown by the dotted, solid, and dashed lines, respectively ($T_{\rm cr}$ =
1800K throughout). b) Predicted colors based on the UCMs with  $T_{\rm cr}$ 
= $T_{\rm cond}$ (case C), 1900, 1800, and 
1700\,K  are shown by the lines labeled with 1, 2, 3, and 4, respectively.
(log\,$g$ = 5.0, $v_{\rm micro}=$1 km s$^{-1}$ and the solar metallicity 
throughout). 
}
\label{Fig1}
\end{figure}

\begin{figure}
\epsscale{0.70}
\vspace{-5mm}
\plotone{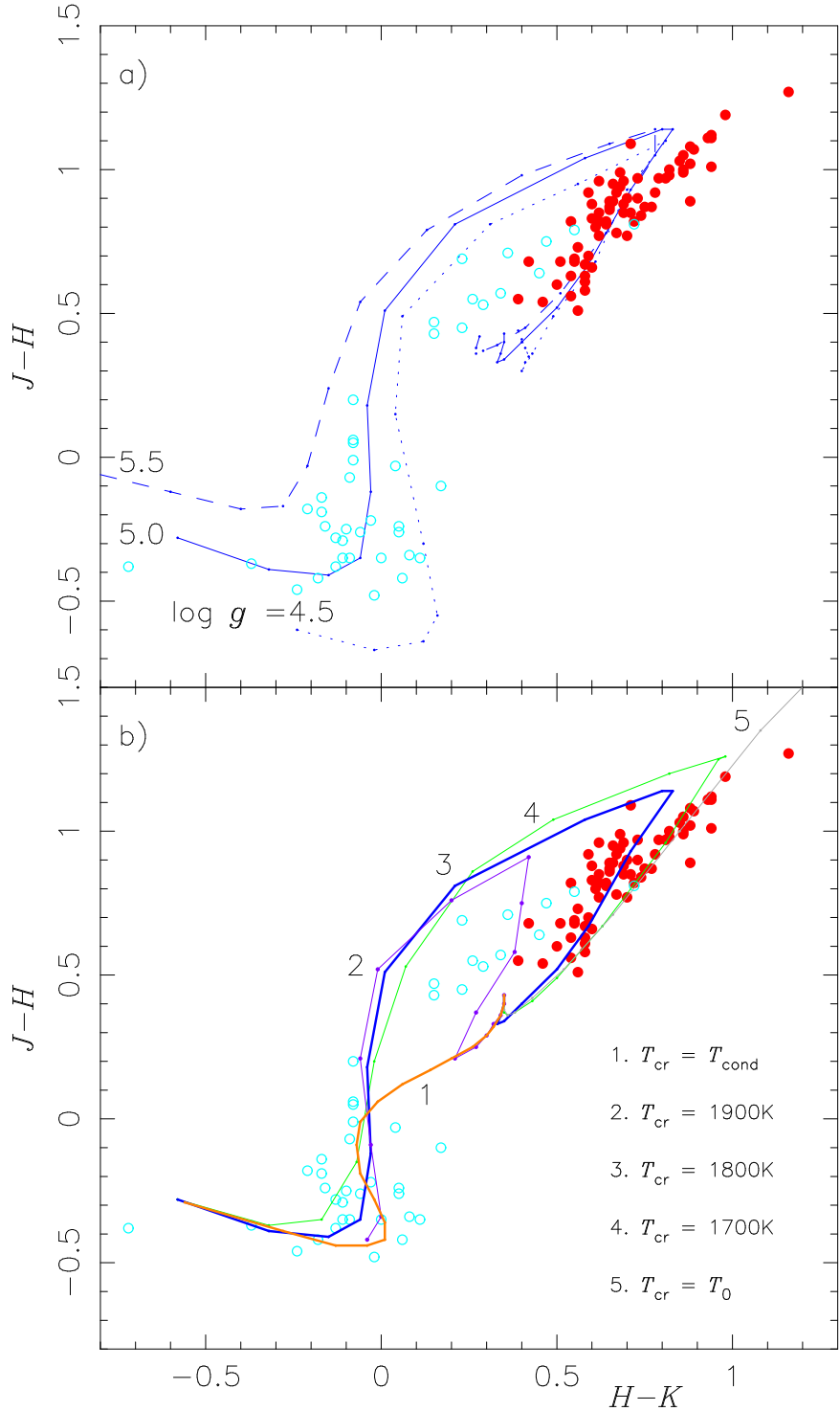}
\vspace{-5mm}
\caption {
Observed ${J-H}$ vs. $H-K$ on the MKO system (Knapp et al.2004)
are shown by the filled and open circles for L and T dwarfs, respectively.
a) Predicted ${J-H}$ vs. $H-K$  based on the UCMs with  log\,$g$ = 4.5, 5.0,
and 5.5 are shown by the dotted, solid, and dashed lines, respectively,
($T_{\rm eff} = 700 - 2600$\,K and $T_{\rm cr} = 1800$\,K throughout). 
b) Predicted ${J-H}$ vs. $H-K$  based on the UCMs with  $T_{\rm cr} = 
T_{\rm cond}$ (case C), 1900\,K, 1800\,K, 1700\,K, and $T_{\rm 0}$ (case B) 
are shown by the lines labeled with 1, 2, 3, 4, and 5, respectively.
($T_{\rm eff} = 700 - 2600$\,K and log\,$g$ = 5.0  throughout). 
}
\label{Fig2}
\end{figure}

\begin{figure}
\vspace{-5mm}
\epsscale{0.50}
\plotone{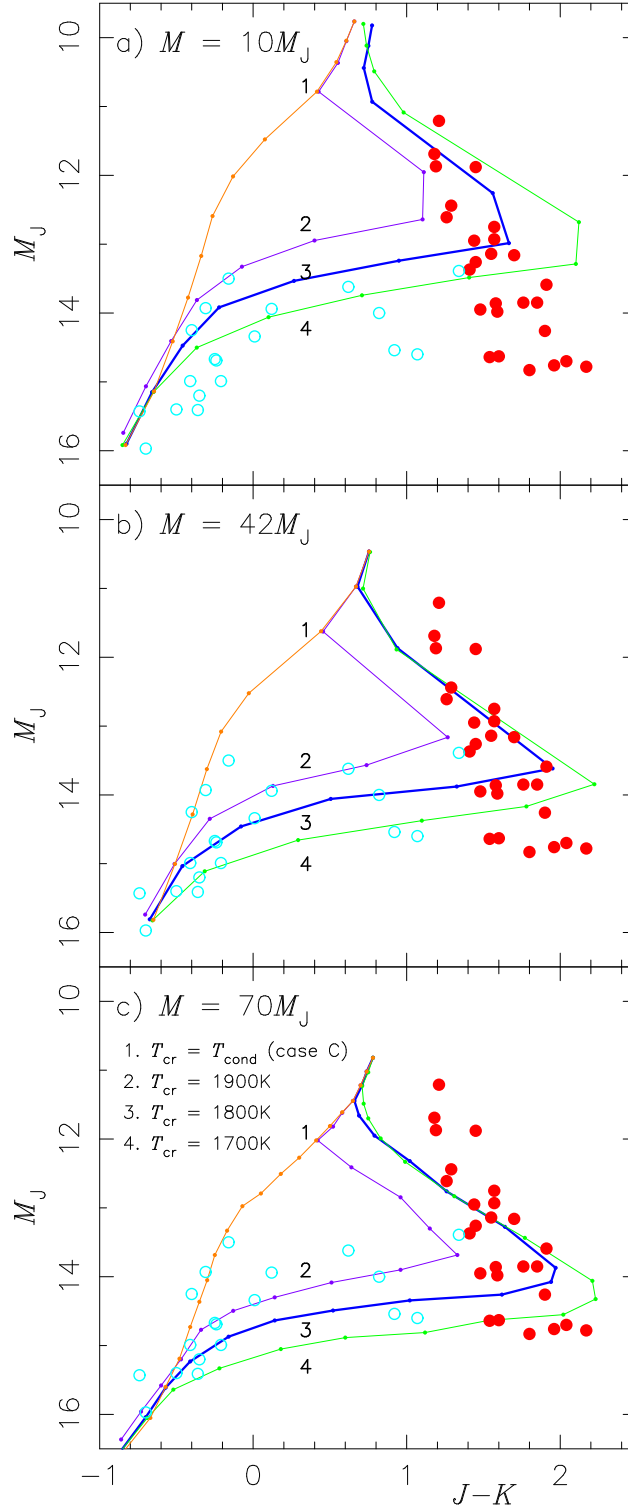}
\vspace{-5mm}
\caption {
Observed $M_{J}$ vs. $J-K$ on the MKO system (Knapp et al.2004)
are shown by the filled and open circles for L and T dwarfs, respectively.
Predicted $M_{J}$ vs. $J-K$  transformed from the evolutionary tracks
by Burrows et al. (1997) via the UCMs with  $T_{\rm cr} = T_{\rm cond}$,
1900\,K, 1800\,K, and 1700\,K  are shown by the  lines labeled with 1, 
2, 3, and 4, respectively.
a) $M = 10 M_{\rm Jupiter}$, b) $M = 42 M_{\rm Jupiter}$, and c) $M = 
70 M_{\rm Jupiter}$.
($T_{\rm eff} = 700 - 2600$\,K and log\,$g$ = 5.0  throughout). 
}
\label{Fig3}
\end{figure}

\begin{figure}
\epsscale{0.8}
\plotone{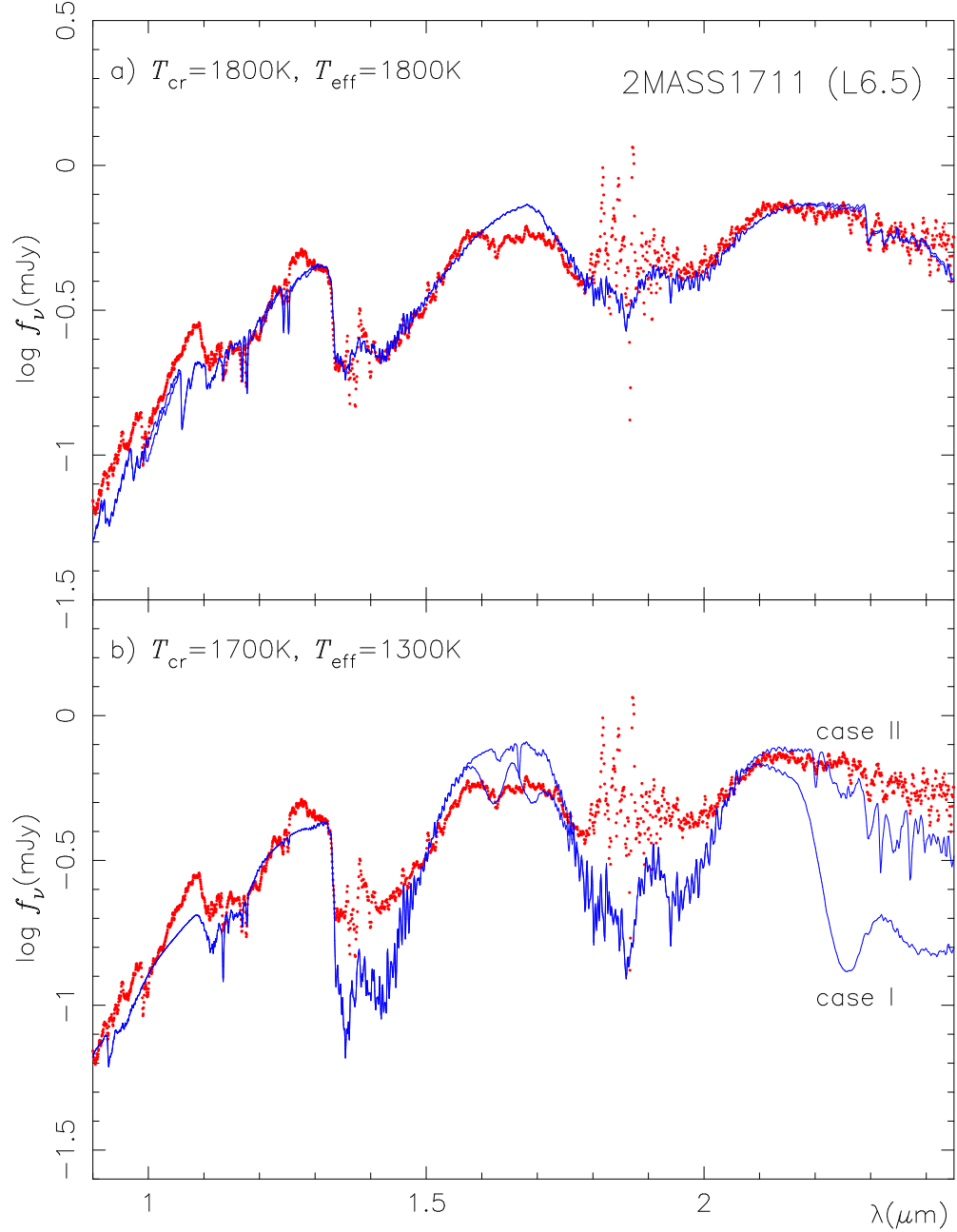}
\caption {
The observed spectrum of 2MASS\,1711 (L6.5) shown by the dots is compared
with the predicted ones based on the UCMs of: a) $T_{\rm cr} = 1800$\,K, 
$T_{\rm eff} = 1800$\,K, and log\,$g$ = 5.0.  
b) $T_{\rm cr} = 1700$\,K, $T_{\rm eff} = 1300$\,K, and log\,$g$ = 5.0.
}
\label{Fig4}
\end{figure}

\begin{figure}
\epsscale{0.8}
\plotone{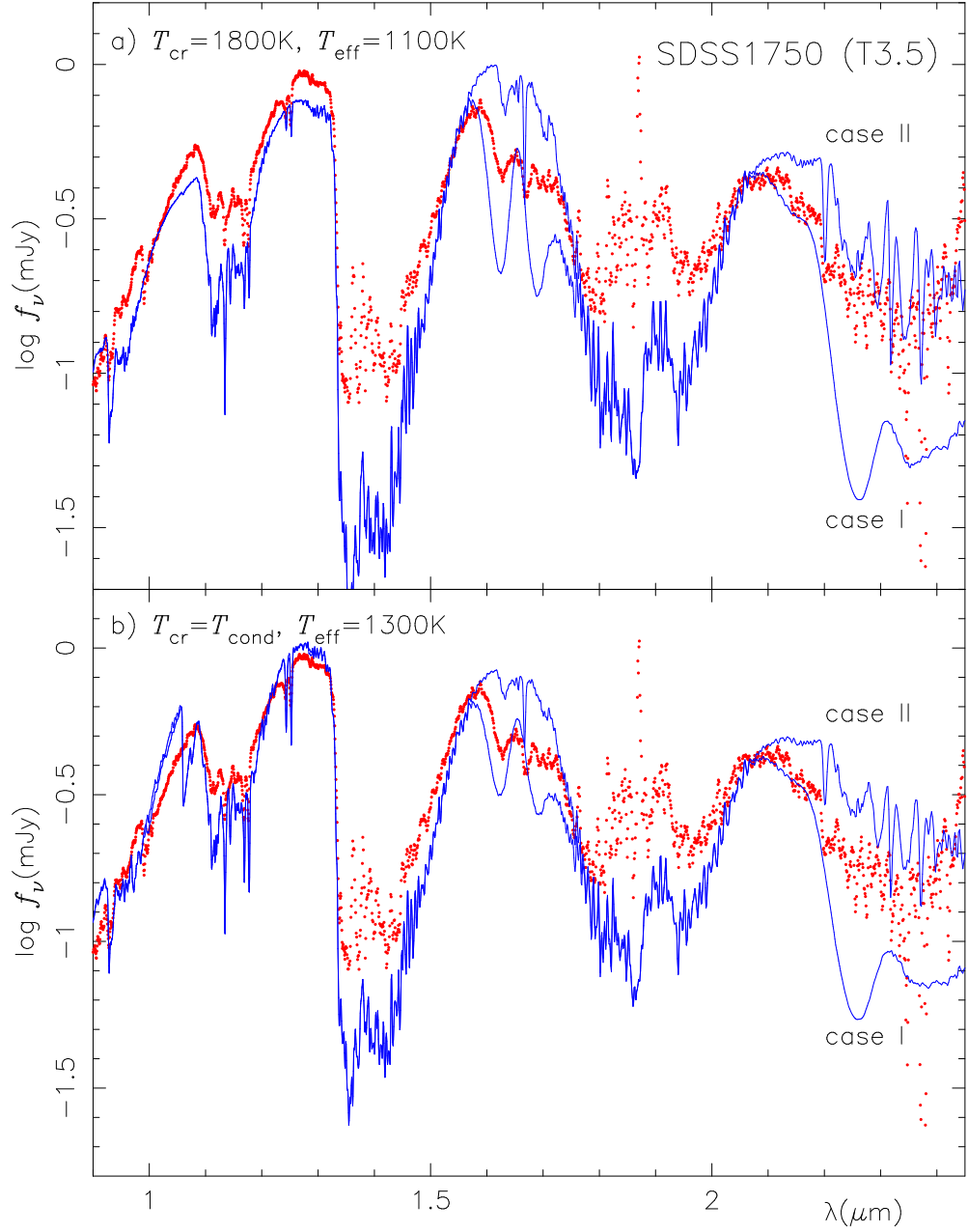}
\caption {
The observed spectrum of SDSS\,1750 (T3.5) shown by the dots is compared
with the predicted ones based on the UCMs of: a) $T_{\rm cr} = 1800$\,K, 
$T_{\rm eff} = 1100$\,K, and log\,$g$ = 5.0.   
b) $T_{\rm cr} = T_{\rm cond}$ (case C), $T_{\rm eff} = 1300$\,K, and 
log\,$g$ = 5.0.
}
\label{Fig5}
\end{figure}

\begin{figure}
\epsscale{0.8}
\plotone{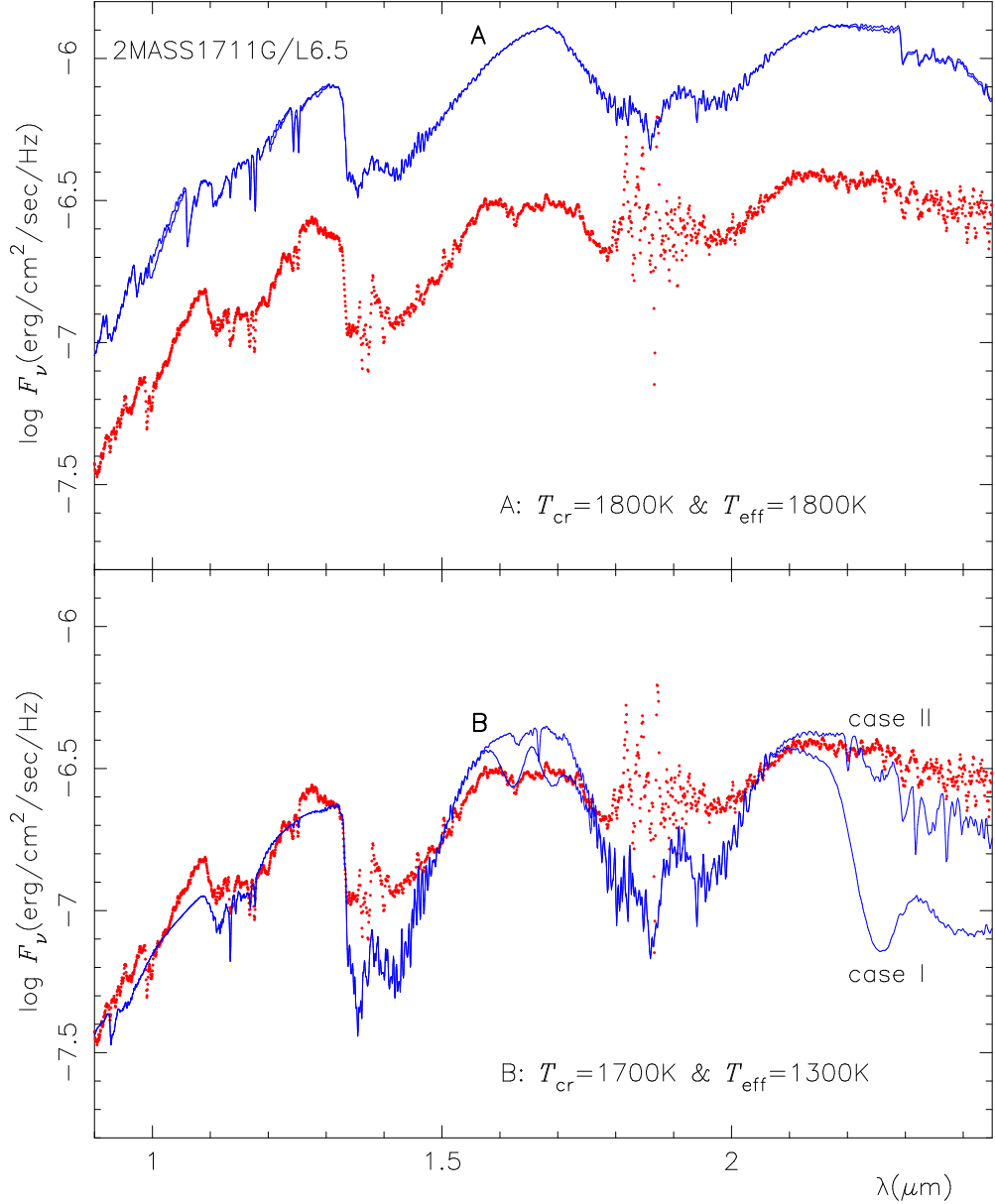}
\caption {
The observed spectrum of 2MASS\,1711 (L6.5) reduced to the emergent
flux on an absolute scale (in unit of erg\,cm$^{-2}$\,sec$^{-1}$\,Hz$^{-1}$)
is shown by the dots. The predicted spectrum based on the UCM
with  $T_{\rm cr} = 1800$\,K, $T_{\rm eff} = 1800$\,K, and log\,$g$ = 5.0  
(cf. Fig.6 of Paper II) is shown by the solid line marked with A . Although 
the observed and predicted spectra can be fitted on the relative scale 
(i.e. by the shapes of the spectra as in Fig.4), they
cannot be fitted on the absolute scale. Infrared colors suggest a lower 
$T_{\rm cr}$ for this object(Table 1; Fig.\,1b), and the predicted
spectrum based on the UCM with  $T_{\rm cr} = 1700$\,K, $T_{\rm eff} = 
1300$\,K, and log\,$g$ = 5.0  is shown by the solid line marked with B.
The observed and predicted spectra can be fitted on the absolute scale,
but the predicted water bands appear to be too strong. 
}
\label{Fig6}
\end{figure}

\begin{figure}
\epsscale{0.8}
\plotone{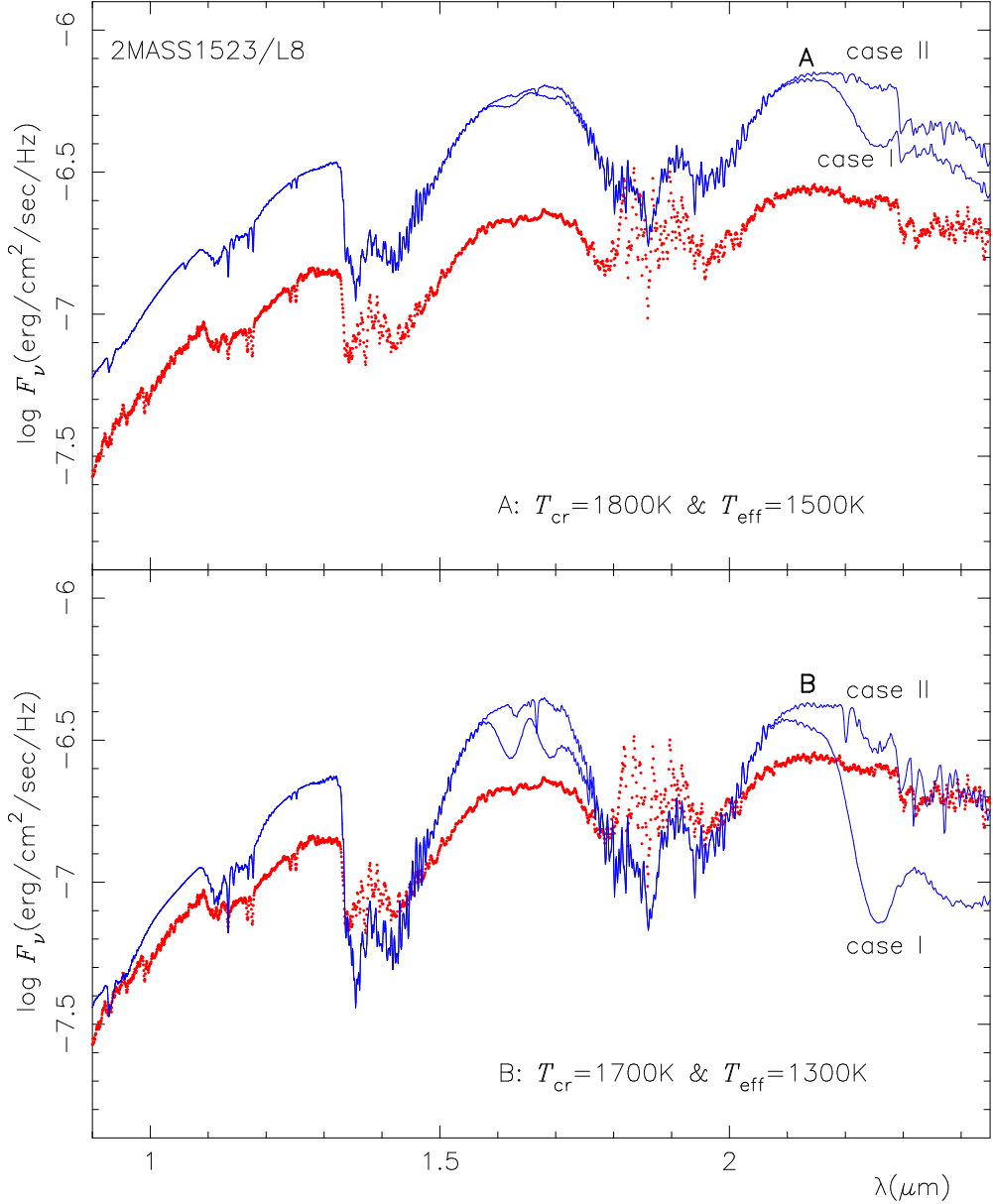}
\caption {
The observed spectrum of 2MASS\,1523 (L8) reduced to the emergent
flux on an absolute scale is shown by the dots. 
The predicted spectrum based on the UCM with  $T_{\rm cr} = 1800$\,K, 
$T_{\rm eff} = 1500$\,K, and log\,$g$ = 5.0  
is shown by the solid line marked with A. Although the observed and
predicted spectra can be fitted on the relative scale (Fig.8 of Paper II), 
it is clear that they cannot be fitted on the absolute scale. Infrared colors 
suggest a lower $T_{\rm cr}$  for this object (Table 1; Fig.\,1b), and the 
predicted
spectrum based on the UCM with  $T_{\rm cr} = 1700$\,K, $T_{\rm eff} = 
1300$\,K, and log\,$g$ = 5.0  is shown by the solid line marked with B.
The observed and predicted spectra cannot be fitted on the absolute scale
as well, but it is to be noted that the uncertainty of the absolute scale 
is about 0.2 - 0.3\,dex.  
}
\label{Fig7}
\end{figure}

\begin{figure}
\epsscale{0.8}
\plotone{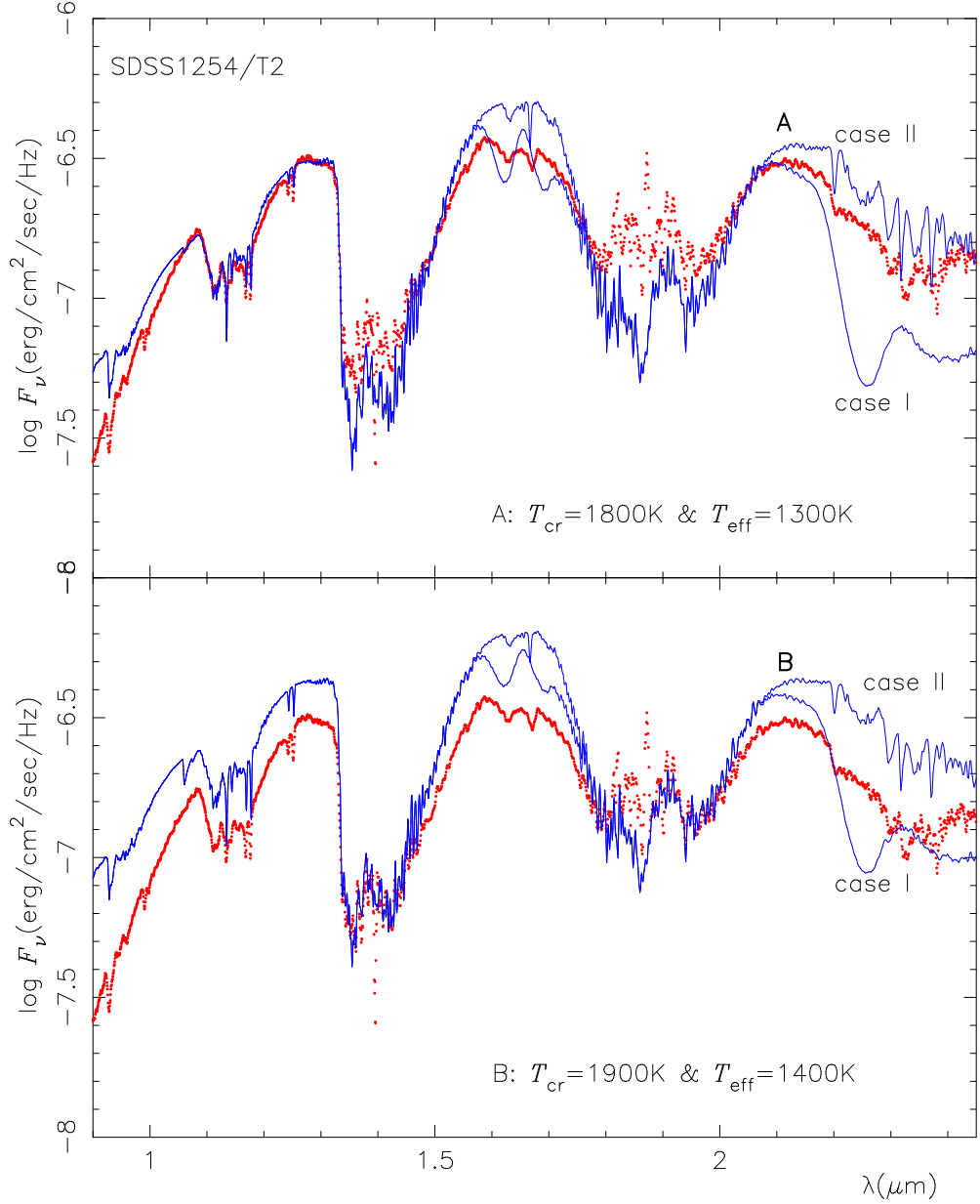}
\caption {
The observed spectrum of SDSS\,1254 (T2) reduced to the emergent
flux on an absolute scale is shown by the dots. 
The predicted spectrum based on the UCM with  $T_{\rm cr} = 1800$\,K, 
$T_{\rm eff} = 1300$\,K, and log\,$g$ = 5.0 is 
shown by the solid line marked with A. The observed and
predicted spectra agree well both on the relative and absolute scales.
Infrared colors suggest $T_{\rm cr} \approx 1800$\,K  for this object(Table 1;
Fig.\,1b), and this also supports the analysis of Paper II (Fig.9).  For
comparison, the predicted spectrum based on the UCM with  $T_{\rm cr} = 
1900$\,K, $T_{\rm eff} = 1400$\,K, and log\,$g$ = 5.0  is shown by the 
solid line marked with B. The curves A and B show 
quite similar and the degeneracy of $T_{\rm cr}$ and  $T_{\rm eff}$
cannot be removed on the spectra without absolute calibration.
}
\label{Fig8}
\end{figure}

\begin{figure}
\epsscale{0.8}
\plotone{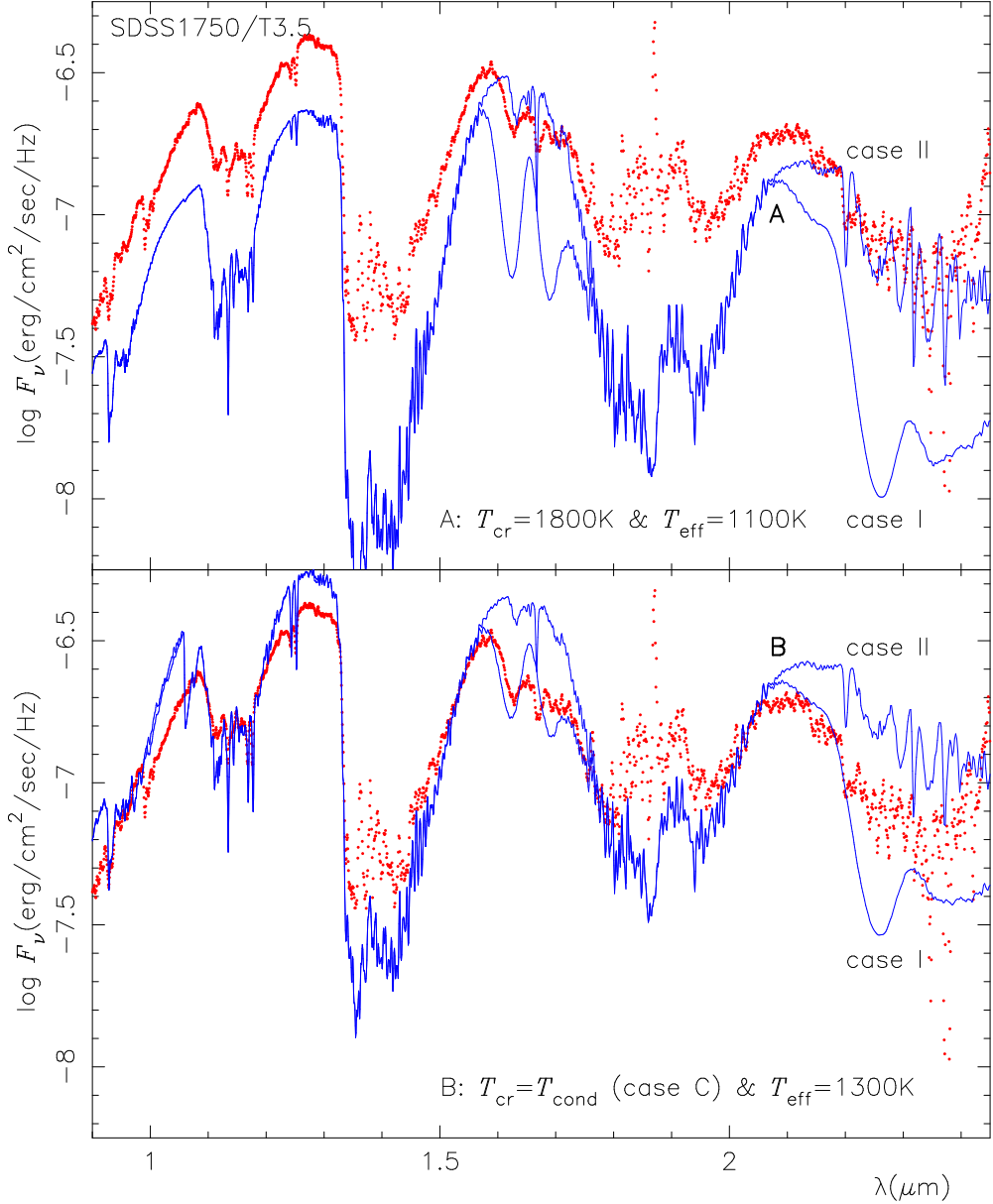}
\caption {
The observed spectrum of SDSS\,1750 (T3.5) reduced to the emergent
flux on an absolute scale is shown by the dots. 
The predicted spectrum based on the UCM
with  $T_{\rm cr} = 1800$\,K, $T_{\rm eff} = 1100$\,K, and log\,$g$ = 5.0  
is shown by the solid line marked with A. The observed and predicted spectra 
agree marginally within the possible uncertainty of the absolute scale,
although the predicted water  bands appear to be too strong (Fig.10 of Paper 
II).  Infrared colors suggest $T_{\rm cr} \approx T_{\rm cond}$ (case C) for 
this object (Table 1; Fig.\,1b) , and
the predicted spectrum based on the UCM with  $T_{\rm cr} = T_{\rm cond}$ 
(case C), $T_{\rm eff} = 1300$\,K, and log\,$g$ = 5.0 is shown by the 
solid line marked with B. The observed and predicted spectra now
show better agreement both on the absolute and relative scales.
}
\label{Fig9}
\end{figure}

\begin{figure}
\vspace{-5mm}
\epsscale{0.60}
\plotone{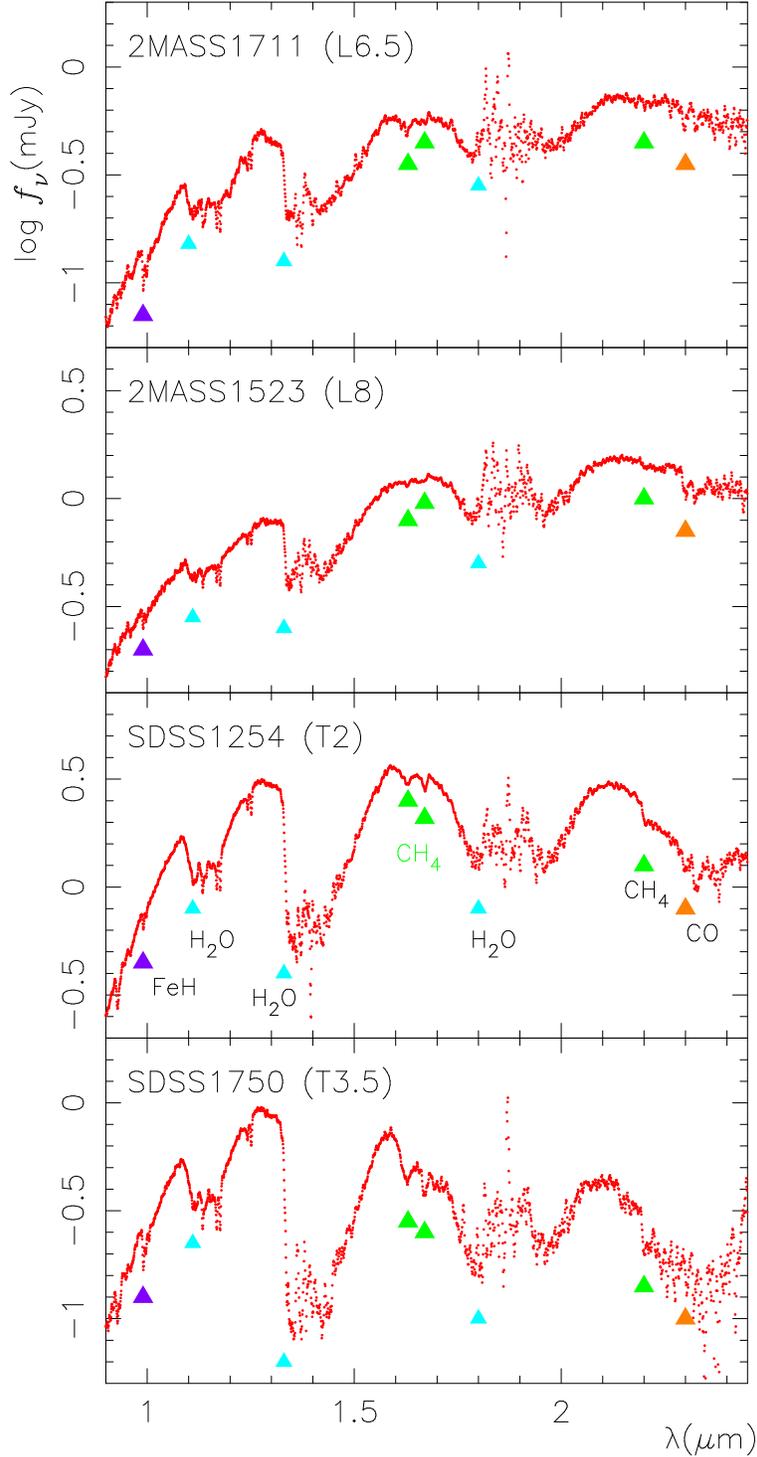}
\vspace{-5mm}
\caption {
The observed spectral sequence between L6.5 and T3.5 are summarized.
These objects 
were interpreted as a temperature sequence ($T_{\rm eff}$ from 1800\,K
to 1100\,K) based on the UCMs with an uniform value of $T_{\rm cr} = 1800$\,K 
(Paper II). The same observed data reduced to an absolute scale
are reanalyzed again with the UCMs, but freed from the assumption of
the uniform value of $T_{\rm cr}$, and the result revealed that the 
values of $T_{\rm eff}$ for all these objects are about the same at 1300\,K
while the values of $T_{\rm cr}$ should increase (or the thickness of
the dust cloud decreases) from L to T dwarfs (see Table 1).
}
\label{Fig10}
\end{figure}

\begin{figure}
\epsscale{0.55}
\plotone{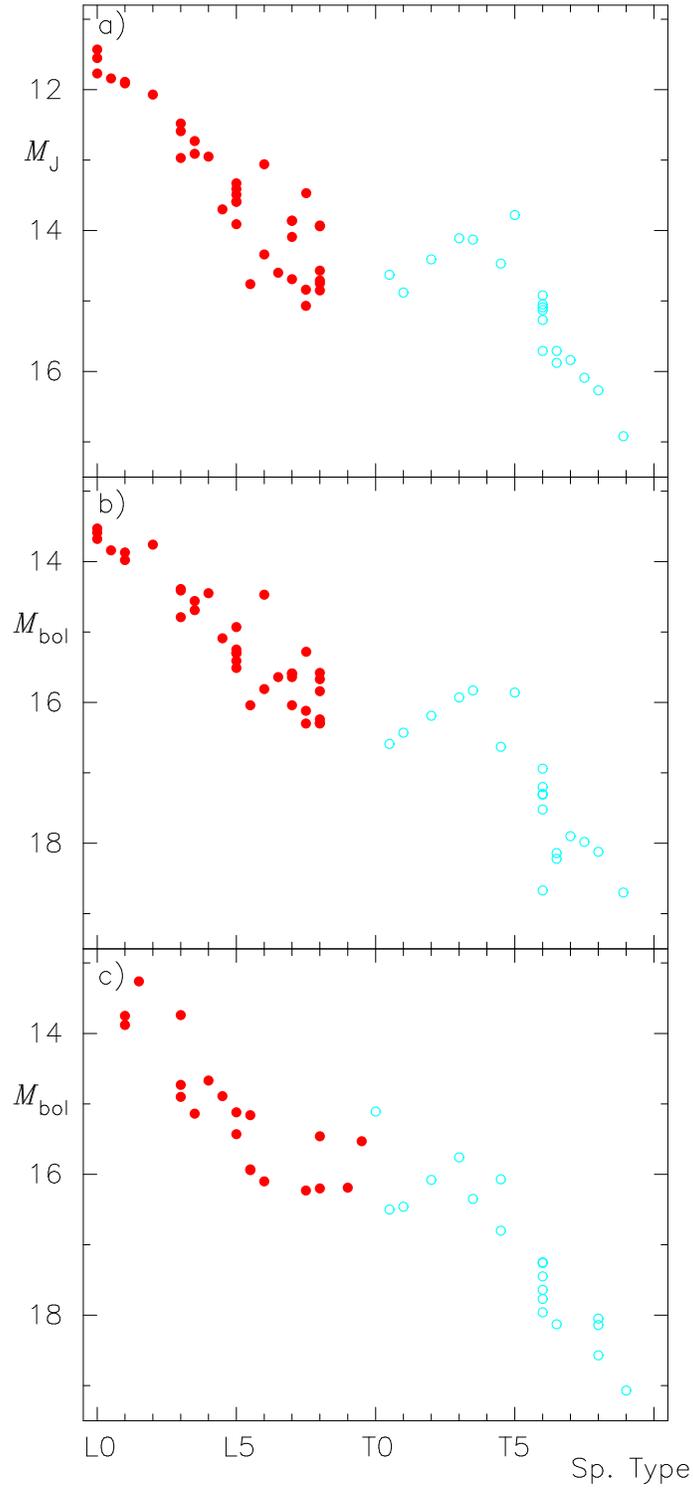}
\caption {
a) $M_{J}$ plotted against L-T spectral types based on the data by Vrba et al.
(2004). 
b) $M_{\rm bol}$ plotted against L-T spectral types based on the data by 
Vrba et al. (2004). 
c) The same as b) but based on the data by Golimowski et al. (2004). 
}
\label{Fig11}
\end{figure}

\begin{figure}
\epsscale{0.55}
\vspace{-5mm}
\plotone{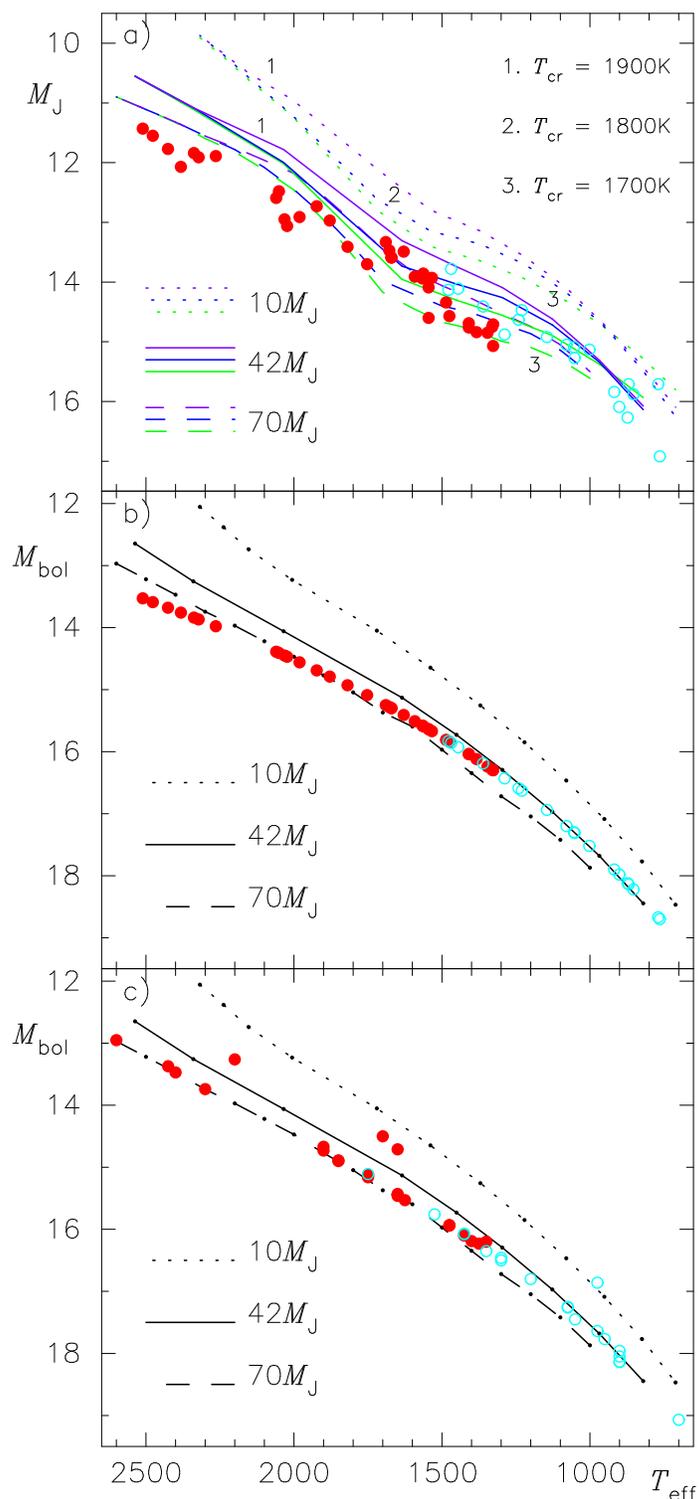}
\vspace{-5mm}
\caption {
a) $M_{J}$ plotted against $T_{\rm eff}$ based on the data by Vrba et al.
(2004). Also $M_{J}$ transformed from $M_{\rm bol}$ of Burrows et al. (1997)
models of $M = 10, 42,$ and $70\,M_{\rm Jupiter}$ via UCMs are
overlaid.  Note that each case is triplicated for  $T_{\rm cr}$ 
=  1900, 1800, and 1700\,K.
b) $M_{\rm bol}$ plotted against $T_{\rm eff}$ based on the data by Vrba et al.
(2004). The evolutionary tracks for $M = 10, 42,$ and 70\,$M_{\rm Jupiter}$ 
by Burrows et al.(1997) are overlaid.  
c) The same as b) but  based on the data  by Golimowski  et al.
(2004) for an age of 3\,Gyr.
}
\label{Fig12}
\end{figure}

\end{document}